\documentclass[iop,numberedappendix,twocolappendix]{emulateapj}
\usepackage{natbib}

\shorttitle{Ultra-long GRBs}
\shortauthors{Virgili et al.}

\newcommand{\swift}{\textit{Swift}}
\newcommand{\fermi}{\textit{Fermi}}

\begin{document}

\title{GRB 091024A and the nature of ultra-long gamma-ray bursts}

\author{F. J. Virgili$^1$, C. G. Mundell$^1$, V. Pal'shin$^2$, C. Guidorzi$^3$, R. Margutti$^4$, A. Melandri$^5$, R. Harrison$^1$, S. Kobayashi$^1$, R. Chornock$^4$, A. Henden$^6$, A. C. Updike$^7$, S. B. Cenko$^{8,9}$, N. R. Tanvir$^{10}$, I. A. Steele$^1$, A. Cucchiara$^{11}$, A. Gomboc$^{12}$, A. Levan$^{13}$, Z. Cano$^{14}$, C. J. Mottram$^1$, N. R. Clay$^1$, D. Bersier$^1$, D. Kopa$\check{\rm c}$$^{12}$, J. Japelj$^{12}$, A. V. Filippenko$^8$, W. Li$^{8,15}$, D. Svinkin$^2$, S. Golenetskii$^2$, D. H. Hartmann$^{16}$, P. A. Milne$^{17}$, G. Williams$^{17}$, P. T. O'Brien$^{10}$, D. B. Fox$^{18}$, and E. Berger$^4$}
\affil{$^1$Astrophysics Research Institute, Liverpool John Moores University, Liverpool, L3 5RF, UK;  F.J.Virgili@ljmu.ac.uk \\ 
$^2$Ioffe Physical Technical Institute, St. Petersburg 194021, Russian Federation \\
$^3$Department of Physics and Earth Sciences, University of Ferrara, Via Saragat, 1, 44122 Ferrara, Italy \\
$^4$Harvard-Smithsonian Center for Astrophysics, 60 Garden Street, Cambridge, MA, 02138, USA \\
$^5$INAF/Brera Astronomical Observatory, via Bianchi 46, 23807, Merate (LC), Italy \\
$^6$AAVSO, 49 Bay State Road, Cambridge, MA 02138, USA \\
$^7$Department of Chemistry and Physics, Roger Williams University, Bristol, RI 02809, USA \\
$^8$Department of Astronomy, University of California, Berkeley, CA 94720-3411, USA \\
$^9$Astrophysics Science Division, NASA Goddard Space Flight Center, Mail Code 661, Greenbelt, MD, 20771, USA \\
$^{10}$Department of Physics and Astronomy, University of Leicester, University Road, Leicester LE1 7RH, UK \\
$^{11}$Department of Astronomy and Astrophysics, UCO/Lick Observatory, University of California, 1156 High Street, Santa Cruz, CA 95064, USA \\
$^{12}$Faculty of Mathematics and Physics, University of Ljubljana, Jadranska 19, 1000 Ljubljana, Slovenia \\
$^{13}$Department of Physics, University of Warwick, Coventry CV4 7AL, UK \\
$^{14}$ Centre for Astrophysics and Cosmology, Science Institute, University of Iceland, Reykjavik, Iceland, 107 \\
$^{15}$Deceased 11 December 2011. \\
$^{16}$Department of Physics and Astronomy, 118 Kinard Laboratory, Clemson University, Clemson, South Carolina 29631-0978, USA \\
$^{17}$MMT Observatory, University of Arizona, Tucson, AZ 85719, USA \\
$^{18}$Department of Astronomy and Astrophysics, The Pennsylvania State University, 525 Davey Lab, University Park, PA 16802, USA \\}

\begin{abstract}

We present a broadband study of gamma-ray burst (GRB) 091024A within the context of other ultra-long-duration GRBs. An unusually long burst detected by Konus-Wind, \swift, and \fermi, GRB 091024A has prompt emission episodes covering $\sim 1300$\,s, accompanied by bright and highly structured optical emission captured by various rapid-response facilities, including the 2-m autonomous robotic Faulkes North and Liverpool Telescopes, KAIT, S-LOTIS, and SRO. We also observed the burst with 8- and 10-m class telescopes and determine the redshift to be $z = 1.0924 \pm 0.0004$. We find no correlation between the optical and $\gamma$-ray peaks and interpret the optical light curve as being of external origin, caused by the reverse and forward shock of a highly magnetized jet ($R_B \approx 100$--200). Low-level emission is detected throughout the near-background quiescent period between the first two emission episodes of the Konus-Wind data, suggesting continued central-engine activity; we discuss the implications of this ongoing emission and its impact on the afterglow evolution and predictions.  We summarize the varied sample of historical GRBs with exceptionally long durations in gamma-rays ($\gtrsim$ 1000\,s) and discuss the likelihood of these events being from a separate population; we suggest ultra-long GRBs represent the tail of the duration distribution of the long GRB population.
\end{abstract}


\section{Introduction}

Following the first detection of a gamma-ray burst (GRB) by a military satellite in the late 1960s (Klebesadel, Strong, \& Olson 1973), the BATSE $\gamma$-ray detector \cite{fishman89} onboard the \textit{Compton Gamma-Ray Observatory (CGRO)} revolutionized the study of $\gamma$-ray properties, detecting flashes with durations from $t<64$\,ms to $t>500$\,s,  showing their sky distribution to be isotropic, and producing a catalog of 1637 GRB light curves (revised 4B catalog; Paciesas et al. 1999). Most notably, GRB $t_{90}$ duration --- defined as the time in which 5\% to 95\% of the burst fluence is accumulated --- has played a key role in GRB classification \citep{kouveliotou93}.  Initially seen as a powerful discriminator between possible GRB progenitor models, $t_{90}$ has been shown to be sensitive to detector energy range \citep{sakamoto11,virgili12,qin13}, thus requiring a more complete approach to progenitor categorization (e.g., Zhang et al. 2009) and the study of emission mechanisms. 

GRB 091024A falls into a category of bursts with observed $\gamma$-ray emission lasting longer than $\sim 1000$\,s.  Such emission may be continuous or consist of shorter ($\sim 50$--250\,s) episodes separated by significant periods of low-level emission or quiescence, complicating the definition of ``duration'' and the interpretation of its physical significance for different bursts.  In some cases, long-lasting emission has been attributed to ongoing central-engine activity.   Observational evidence to support this hypothesis has been seen at longer wavelengths in the form of X-ray \cite{burrows05,zhang06,fan06,chincarini10,margutti11} and optical flares \citep{vestrand05,boer06,wei06,melandri09} that show characteristics, such as  short-timescale variability, steep rise and decay slopes, and a clear lag-luminosity relation \citep{margutti10}, that are, in some cases, difficult to reconcile within the standard fireball model or an external-shock origin  (Melandri et al. 2010; Kopa$\check{\rm c}$ et al. 2013).  Further support comes from long-lasting X-ray emission prior to the steep decay phase of the X-ray light curve \citep{zhang06}, interpreted as curvature radiation from the cessation of central-engine activity \citep{zhang06,liang06b,yamazaki06,zhangbb09}, and long-duration X-ray-rich bursts \citep{feroci01,nicastro04,intZand03}.  

A small number of BATSE bursts were detected with  prompt emission lasting $> 500$\,s and up to 1300\,s, but their poor localizations and the resultant lack of multi-wavelength counterparts limit understanding of their nature and origin. 
The launch of NASA's \textit{Swift} satellite \citep{gehrels04} with its optimized GRB detection and rapid dissemination of accurate localizations, coupled with real-time follow-up observations by autonomous robotic optical telescopes such as the 2-m Liverpool and Faulkes telescopes \cite{guidorzi06} and smaller very rapid-response facilities such as KAIT\footnote{http://astro.berkeley.edu/bait/public\_html/kait.html .} (Filippenko et al. 2001; Li et al. 2003), Super-LOTIS \citep{park97,park02}, and SRO\footnote{http://www.aavso.org/sonoita-research-observatory-sro .}, has opened a new era of multi-wavelength study of GRB properties at early times.

Although ultra-long GRBs remain rare, detection and comprehensive follow-up observations of ultra-long events such as GRBs 091024A, 110709B \citep{zhang12}, and 111209A \citep{gendre12,stratta13} are providing new insights into the physics of this extreme subset.  These bursts are an interesting laboratory in which to test the framework of the internal/external-shock model and our assumptions of central-engine activity, most notably in the context of accretion onto a black hole from a very large star (e.g., Gendre et al. 2013; Levan et al 2013, Stratta et al. 2013). 

In this paper, we present a detailed analysis of the $\gamma$-ray, X-ray, and optical emission from  GRB~091024A, whose observed prompt emission lasted for $\gtrsim 1200$\,s, allowing simultaneous multi-wavelength observations to be obtained. We show that the optical light curve is consistent with an external-shock origin, and that there is significant $\gamma$-ray emission detected in the period of apparent quiescence between the first two episodes of emission, which has a measurable impact on the observed optical light curve at early times. We place GRB~091024A into a wider context by comparing its properties with a sample of ultra-long GRBs with duration $\gtrsim$ 1000\,s and discuss whether they represent a new emerging class of GRBs. 

The multi-wavelength observations of GRB~091024A are presented in \S 2, and the temporal and spectral analysis together with the derived energetics are presented  in \S 3.  Section 4 is dedicated to modeling the afterglow. In \S 5 we introduce the sample of historic ultra-long GRBs, discussing their individual properties as members of two classes --- continuous and intermittent prompt $\gamma$-ray emission.  Section 6 presents a discussion and \S 7 our conclusions. Details of each individual burst in our sample are summarized in the Appendix.  Throughout the paper we use the following conventions: UT dates are used and are relative to the \swift/BAT trigger at $T_0 = 08:56:01$ on October 24, 2009; $F(\nu, t) \propto t^{-\alpha} \nu^{-\beta}$;  The following cosmological model is used: $H_0 = 71$\,km\,s$^{-1}$\,Mpc, $\Omega_m = 0.3$, $\Omega_\Lambda = 0.7$; Uncertainties are quoted at 1$\sigma$ unless otherwise specified.
 
\section{Observations}

\subsection{Gamma-ray and X-ray Observations}

\begin{figure}
\includegraphics[angle=0,scale=0.53]{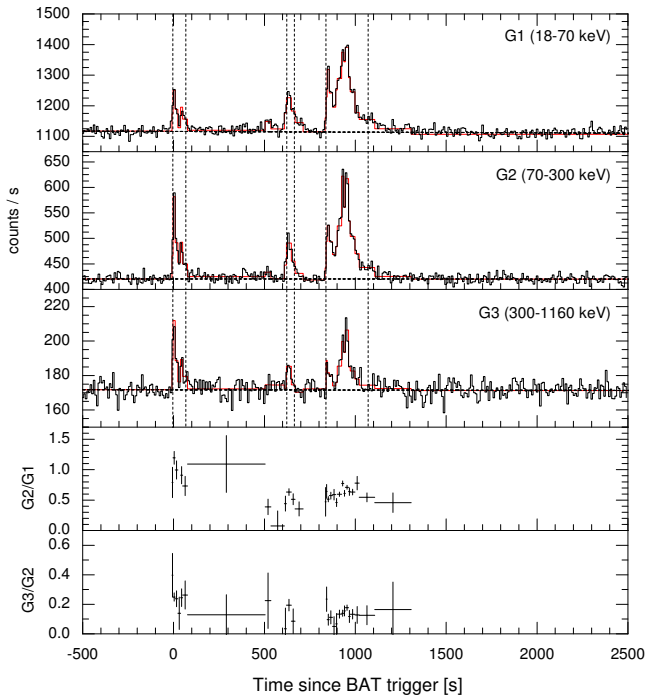}
\caption{Hardness ratios and light curves of GRB 091024A in the three Konus-Wind bands.  Dashed lines are approximate \fermi/GBM durations, highlighting the slight underestimation of the durations of the second and third emission episodes.}
\label{kw_lc}
\end{figure}

GRB 091024A was detected at high energies by Konus-Wind  \citep[KW;][]{golenetskii09}, \swift~BAT \cite{marshall09} and XRT \cite{page09}, \fermi -GBM \cite{bissaldi09}, and SPI-ACS \cite{gruber11}.  The data from  KW and GBM cover the entire burst duration, with GBM having triggered a second time on the major outburst about 600\,s after the first trigger \cite{gruber11}.  The KW light curve in Figure \ref{kw_lc} shows three multi-peak emission episodes: the first with duration $\delta t \approx 88$\,s at $T+T_{0,KW}=-8.9$\,s, the second with $\delta t \approx 106$\,s at $T+T_{0,KW} = 609$\,s, and the third with $\delta t \approx 477$\,s at $T+T_{0,KW}= 883$\,s (with respect to the KW trigger time; \swift\ $T_0 -$ KW $T_0 \approx 0.3$\,s). 
 
\begin{figure}
\includegraphics[angle=0,scale=0.9]{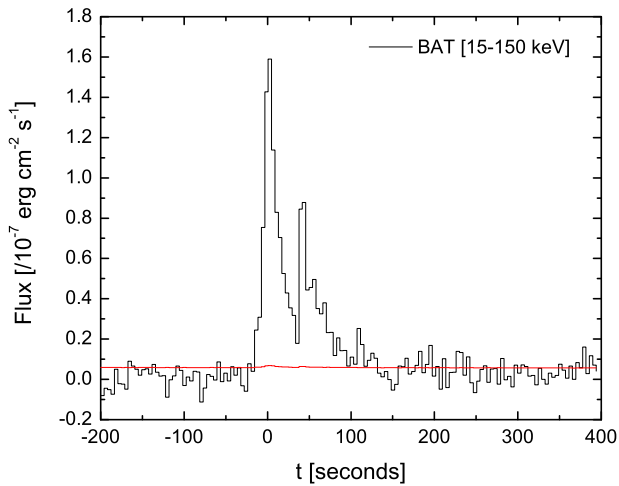}
\caption{\swift-BAT 4\,s binned light curves (15--150\,keV) of the first emission episode  ($T_0 < 450$\,s) of GRB 091024A.  The remaining $\gamma$-ray emission was observed but no coded-mask information is available.}
\label{swift_lc_a}
\end{figure}

\begin{figure}
\includegraphics[angle=0,scale=0.45]{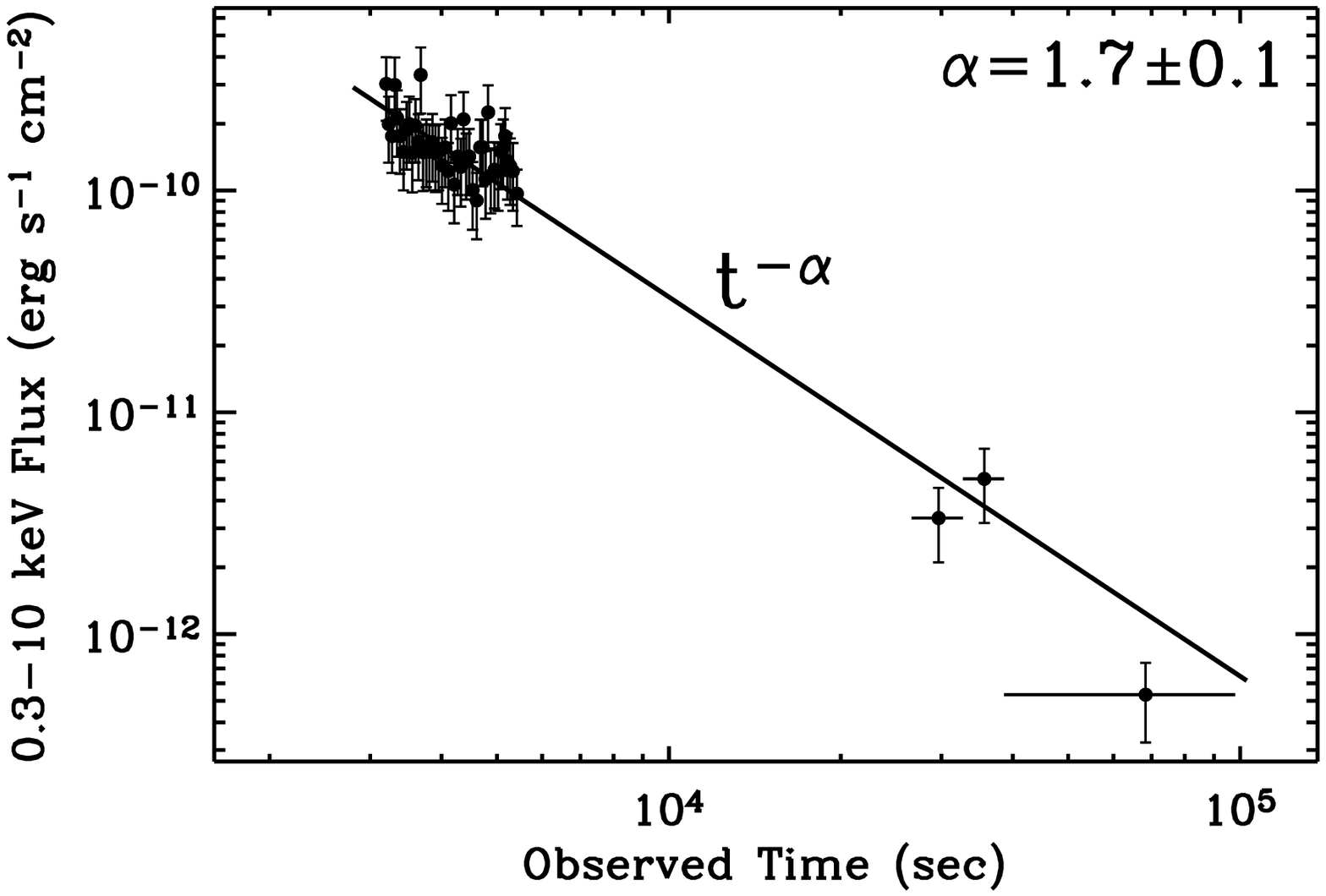}
\caption{\swift-XRT light curve with a simple power law fit overlaid ($\alpha = 1.7 \pm 0.1$). Observations began at $T = T_0+53$\,min.}
\label{swift_lc_b}
\end{figure}

\textit{Swift}-BAT and XRT have partial datasets which are truncated because of Earth-limb constraints. The former contains the first emission episode (Figure \ref{swift_lc_a}) and the latter emission from $T_0+53$ min \cite{page09} to $T_0+1398$ min (Figure \ref{swift_lc_b}).   We processed the BAT data with standard HEAsoft tools (v 6.10) and utilize the spectra for a joint analysis with the first peak of the KW data.  

Although truncated, the XRT X-ray light curve can be adequately fit with a simple power law having a decay slope $\alpha = 1.7 \pm 0.1$, following the procedure of Margutti et al. (2013a).  The X-ray spectrum can be fit with a simple power law of slope $\Gamma_X = 1.49^{+0.23}_{-0.21}$, with total column density $N_{\rm H} = 1.7^{+1.2}_{-1.1} \times 10^{22}$\,cm$^{-2}$ and instrinsic $N_{\rm H} = 3.0^{+1.7}_{-1.5} \times 10^{22}$\,cm$^{-2}$ \citep{evans07,evans09,kalberla05}. 
 
\subsection{Optical Observations}

A number of rapid-response optical facilities, with apertures ranging from 0.35-m to 2-m, responded to the {\it Swift}-BAT trigger beginning at $T_0+58$\,s. Optical monitoring continued to $T_0+10^6$\,s, along with redshift determinations by 8- to 10-m class telescopes. We cross-calibrated our photometric dataset with respect to a common set of standard stars observed in $BVRI$ filters with the SRO telescope during the week after the burst event. Final calibrated magnitudes and extinction-corrected fluxes are summarized in Table \ref{phot2} and Figure \ref{optical_gamma}.
 
\subsubsection{Super-LOTIS}

The 0.6-m Super-LOTIS telescope began observing at $T_0+58$\,s for a total of 36\,min \cite{updike09}. We began with a series of $5 \times 10$\,s exposures, then a sequence of $5 \times 20$\,s exposures, and finally increasing to 60\,s exposures after 5\,min. We grouped and coadded several frames during the observing interval in order to increase the total signal-to-noise ratio (S/N), and we calibrated these images with respect to the $R$ band. 

\subsubsection{Katzman Automatic Imaging Telescope}

The 0.76-m Katzman Automatic Imaging Telescope (KAIT) began observations at $T_0+82$\,s \cite{chornock09} with 20\,s exposures in alternating $V$, $I$, and unfiltered bands. Observations ended $\sim 16$\,min after the trigger. $V$-band and $I$-band images have been calibrated with respect to the corresponding filter and unfiltered frames have been calibrated with respect to the $R$ band (Li et al. 2003). 

\subsubsection{Faulkes Telescope North}

The 2-m Faulkes Telescope North (FTN) began monitoring at  $T_0+196$\,s, automatically identifying the optical afterglow \citep{mundell09,cano09}, continuing observations until $T_0+1.2\times10^4$\,s. A series of images in alternating $BVRI$ filters were taken, with exposure times in the range of 10--180\,s.  

\subsubsection{SRO/AAVSO}

The 0.35-m Sonoita Research Observatory Telescope (SRO) began observing at $T_0+540$\,s lasting for about an hour \cite{henden09}.  They observed a series of $V$, $R_C$, and $I_C$ images with exposure times of 180 and 300\,s. Images have been calibrated with respect to the $V$, $R$, and $I$ filters. 
 
\subsubsection{Liverpool Telescope}

The 2-m Liverpool Telescope (LT) provided late-time coverage from $T_0+(4.2-6.4) \times 10^4$\,s.  Five images were taken with exposures of 1800 or 3600\,s, detecting the afterglow in both the $R$ and $I$ bands. 

\subsubsection{Gemini Imaging}

The 8-m Gemini North telescope provided additional late-time coverage with a detection in $r'$ and $i'$ from $5\times180$\,s exposures beginning at $T_0+2.8$\,days and an upper limit in $i'$ from a $9\times200$\,s exposure beginning at $T_0+22.8$ days. Data was reduced using the standard Gemini pipeline tools within the IRAF/gemini package.

\subsection{Optical Light Curve Fitting}

We model the optical light curve with superimposed broken power law components. In order to better characterize the optical behavior we used the best-sampled optical bands ($R$ and $I$). No strong evidence for color change is observed in these bands during the peak episodes in the $10^{2}$--$10^{4}$ \,s time interval.  For that reason we rigidly shift the $I$-band flux to the $R$-band flux (by a factor 0.75) before performing the multi-component fit of the light curve, in order to have the best sampling of the different peaks. The results of the fit are reported in Table \ref{fits} and overlaid with the optical observations in Figure \ref{optical_fitting}. As can be seen, the behavior in the optical band is better described by the sum of three broken power laws ($\chi^{2}_{\rm reduced} = 1.43$, d.o.f. = 75). The final data point obtained by Gemini ($T_0 + 2\times10^6$\,s)  is an upper limit and not used in the model fitting.

If we consider all the optical bands individually we note that evidence of some color change at early times becomes greater. While variation between the $R$ and $I$ bands is compatible with the uncertainties of the combined fit (Table \ref{fits}), there are small variations in the rising slopes ($\alpha$) of individual filters for $t<400$\,s between bluer ($B$ and $V$) and redder ($R$ and $I$) filters: $\alpha_{V} \approx -3.3$, $\alpha_{R} = -2.35 \pm 0.15$, and $\alpha_{I} = -2.53 \pm 0.53$. After the first peak we do not have good sampling of bluer filters and it becomes hard to compare their light curve shapes with the parameters of the composite fitted function reported in Table \ref{fits} for the redder filters. This might indicate some sort of color evolution before and around the time of first peak that is visible only at shorter wavelengths. At late times no color evolution is detected. Observations with simultaneous color information, such as the RINGO3 polarimeter \cite{arnold12}, would greatly improve the availability of color information.

\begin{figure}
\includegraphics[angle=0,scale=0.95]{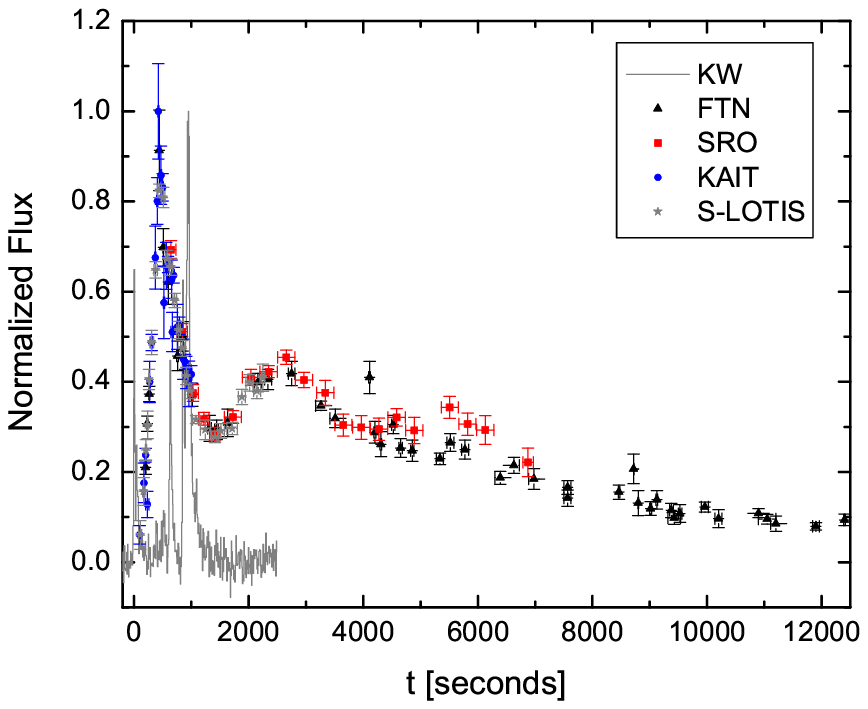}
\includegraphics[angle=0,scale=0.95]{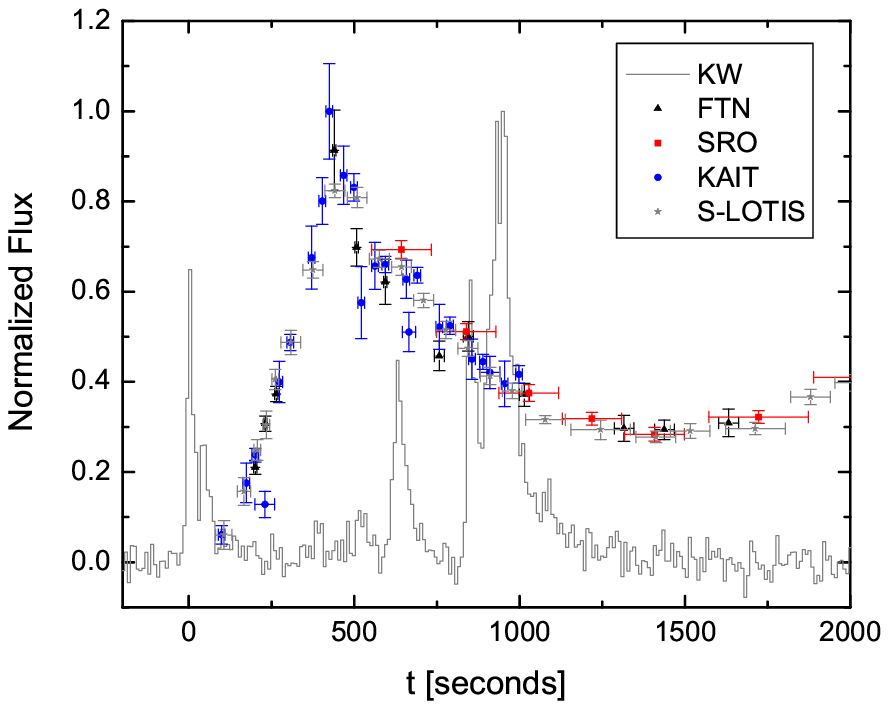}
\caption{Normalized optical (symbols) and KW $\gamma$-ray light curves (18--1160\,keV; gray line) plotted on a linear scale for temporal comparison.  Colors indicate the telescope used. Late-time LT and Gemini data points at $T_0+10^4-10^6$\,s are omitted for clarity.}
\label{optical_gamma}
\end{figure}

\begin{figure}
\includegraphics[angle=0,scale=0.7]{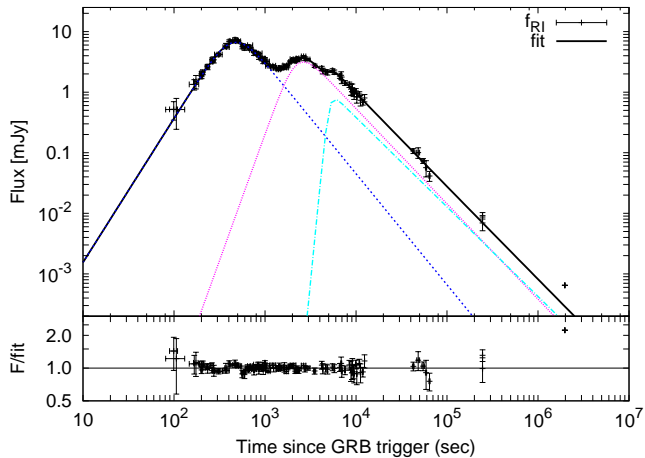}
\caption{Multi-component fit to the $R$ and $I$ (shifted) optical light curve of GRB 091024A. Individual components are in color (dashed) and the total in black (solid).  The last data point is an upper limit and is not used to constrain the model.  Observed photometric data are summarized in Table 1 and model fits are reported in Table 2.}
\label{optical_fitting}
\end{figure}

\subsection{Spectroscopy and Redshift Determination}

\subsubsection{Keck Spectroscopy}

We obtained a spectrum of GRB\,091024A with the Low-Resolution Imaging
Spectrometer (LRIS; Oke et al. 1995) mounted on the 10-m Keck I telescope
beginning at 11:01 on 24 October 2009 ($T_0+2.08$\,hr).  We employed the 5600\,\AA\ dichroic 
beam splitter, with the 400\,lines\,mm$^{-1}$ grism blazed at 3400\,\AA\
on the blue side and the 400\,lines\,mm$^{-1}$ grating blazed at 
8500\,\AA\ on the red side (corresponding to $\sim 7$\,\AA\ resolution
on both ends).  A total of 1200\,s (1250\,s) exposure time was accumulated 
in multiple images for the blue (red) sides, covering a combined 
wavelength range from the atmospheric cutoff to $\sim$ 10,000\,\AA.

All spectra were reduced in the IRAF\footnote{IRAF is distributed by the
National Optical Astronomy Observatory, which is operated by the
Association for Research in Astronomy, Inc., under cooperative agreement with
the National Science Foundation (NSF).} environment using standard
routines.  Cosmic rays were removed using the LA Cosmic routine \citep{v01}.  
Spectra were extracted optimally \citep{h86}, and wavelength calibration was 
performed first relative to Hg-Cd-Zn-Ar lamps and then adjusted slightly 
based on night-sky lines in each individual image.  Both air-to-vacuum and
heliocentric corrections were then applied to all spectra. Flux calibration
was performed by comparison with the spectrum of a spectrophotometric 
standard star. Finally, telluric atmospheric absorption features were 
removed through division by the standard-star spectrum in the relevant
regions \citep{wh88,mfh+00}.

\begin{figure}
\includegraphics[angle=0,scale=0.32]{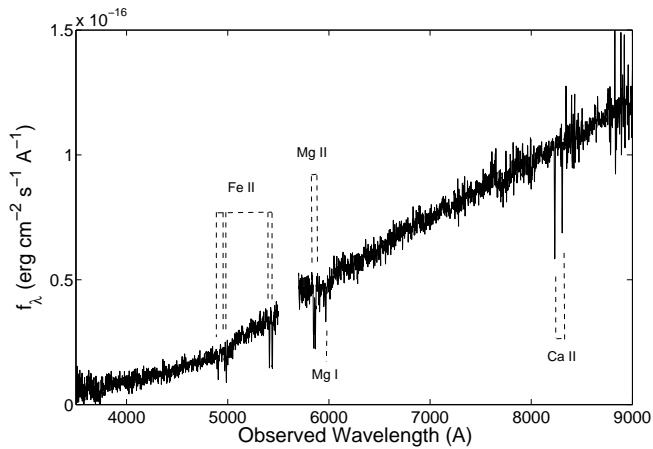}
\caption{Spectrum of GRB 091024A obtained with LRIS on the 10-m Keck~I telescope at $T_0+2.08$\,hr.  Strong absorption features and components (Table \ref{lris}) imply $z = 1.0924 \pm 0.0002$.}
\label{keck}
\end{figure}

We identify a series of strong atomic absorption transitions which are
presented in Table~\ref{lris} (see Figure \ref{keck}).  All of the 
detected components are consistent
with a redshift $z = 1.0924 \pm 0.0002$.  The flux below $3500$\,\AA\
is consistent with zero, and the lack of Ly-$\alpha$ absorption redward of this
value limits the host-galaxy redshift of GRB\,091024A to $z \lesssim 1.9$.  

\subsubsection{Gemini Spectroscopy}

Gemini-N equipped with the GMOS camera began to perform
spectroscopic observations at $T_0+2.38$\,hr. The target was visible in the 60\,s $i$-band acquisition image and placed in a $1''$ slit. We used the R400/800 grating configuration which allowed us to cover  the 6000--10,000\,\AA\ wavelength range with a resolution of $R \approx 1200$ at the midpoint.
Two 900\,s  spectra were obtained, followed by a flatfield and a comparison lamp spectrum with 
the same configuration. All of the raw data were processed with the dedicated {\sc GEMINI}
and {\sc GMOS} tools inside the {\sc IRAF} environment. Flatfielding, sky-background subtraction, 
and cosmic-ray rejection were performed, and one-dimensional spectra of the afterglow and the comparison lamp were extracted using the \textit{APALL} task.
We derived the wavelength solution and applied it to our afterglow spectra.

The resulting spectra were coadded to increase the S/N.
In the entire wavelength range spanned by our data we identified Mg~I and Ca~H\&K
absorption features. We fit Voigt profiles to these features, resulting in the following
rest-frame equivalent widths: $W_{{\rm Mg~I}~\lambda 2853} = 1.17 \pm 0.62$, 
$W_{{\rm Ca~II}~\lambda 3934} = 1.38 \pm 0.44$, $W_{{\rm Ca~II}~\lambda 3969} = 1.66 \pm 0.82$.  These strong lines were consistent with a common redshift for the host galaxy of
$z = 1.092$.

\section{Data Analysis}

\subsection{Temporal Analysis}

The most obvious characteristic that sets GRB 091024A apart from other bursts is its extremely long duration, with an episode of $\gamma$-ray emission coincident with the \textit{Swift} trigger as well as two subsequent emission episodes peaking at about $T_0+650$ and 950\,s in both \fermi/GBM and KW.  From the GBM data alone it is difficult to tell whether there is low-level emission during the periods between the emission episodes due to the large fluctuations in the background.  In contrast, the KW light curves have a very flat baseline and, as detailed below, we find significant low-level activity in the long interpulse interval (see Figure \ref{kw_lc}). In addition, we find that the duration of the second emission episode is underestimated in the GBM data, and we perform all of our analyses with the KW derived durations.

\begin{figure}
\includegraphics[angle=0,scale=0.9]{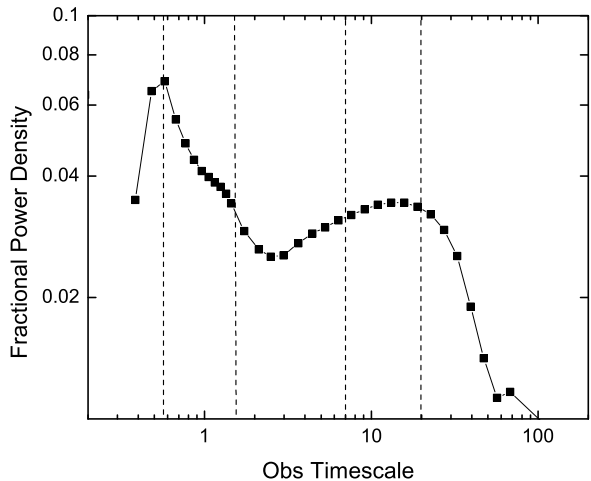}
\caption{Summary of the fractional power density analysis on the entire \swift/BAT non-mask weighted data from $-200 < T_0 < 1200$\,s .  The four timescales of $0.6 \pm 0.2$, $1.4 \pm 0.2$, $7 \pm 0.03$, and $20 \pm 0.03$\,s are indicated with dashed lines.  Errors are estimated by Monte Carlo simulations.}
\label{raf}
\end{figure}

In order to probe the activity of the central engine, we perform a power-spectrum analysis of the unmasked \swift-BAT and KW $\gamma$-ray data in the time domain \cite{li01}.  Specifically, we calculate the fractional power density of the signal for the entire time interval, as well as various temporal epochs and spectral regimes.  This quantity gives a measure of the intrinsic time variability in the signal (see Margutti et al. (2008) and Margutti (2009)\footnote{PhD thesis. Available at http://boa.unimib.it/handle/10281/7465 For details on this method, see specifically Chapter 6. Most inferences based on standard tools like the FFT require stationary signals where the duration is much longer than the typical pulse width, which is not the case for GRBs. Additionally, care must be used in the interpretation of the Fourier spectrum of an aperiodic signal in the time domain, since GRBs are strictly non-periodic (see e.g. Li \& Muraki 2001). The TTD analysis is designed to overcome these limitations and is optimized to study the variability time-scale of short, non-repetitive, non-stationary signals, like GRBs.} for further details) from which we can infer the activity of the central engine powering the GRB.  Using this method, we find two short and two long characteristic timescales at $0.6 \pm 0.2$ and $1.4 \pm 0.2$\,s, and $7 \pm 0.03$ and $20 \pm 0.03$\,s, respectively (See Figure \ref{raf}).  Errors are estimated using Monte Carlo simulations.  

Perhaps more importantly, we find different variability timescales at different temporal epochs during the burst emission.  The first ($T_0$-200 to $T_0$+250\,s) and third ($T_0$+800 to $T_0$+1200\,s) slices include the first and final emission episode, while the second ($T_0$+250 to $T_0$+800\,s) includes the low-level emission period and the weak second emission episode.  The epochs with large amounts of emission show variability at very short timescales, while the second epoch shows only longer variability timescales, behaving similarly to the empirical luminosity-variability relation \cite{fenimore00,reichart01,guidorzi06b}.  The strong similarity in short variability timescales for the first and third episodes suggests that these episodes share a common origin and that the central engine powering the beginning of the burst likely reactivated to produce the later emission.  We also performed this analysis over the individual BAT and KW energy channels.  We find consistency with the trends in the full signal and additionally note that the softest energy channels (15--25\,keV in BAT, 18--70\,keV in KW) show little to no fractional power at low timescales, indicating less intrinsic variability at these energies. 

\subsection{Spectral Analysis}

\subsubsection{Pulse Properties}

We performed both time-integrated and time-resolved analyses on the KW and BAT spectra.  We do not include a joint analysis with \fermi-GBM spectra due to the rapidly changing background caused by the motion of the spacecraft, as discussed above.   First, we consider the spectral parameters of the three emission episodes of the KW data.  The spectra were modeled with cutoff power law and Band (with $\beta = -2.5$) models using the 3-channel data.  For the Band model \cite{band93} with fixed $\beta  = -2.5$ we find $\alpha = (-1.09^{+0.10}_{-0.08}, -1.57^{+1.96}_{-0.17}, -1.46^{+0.06}_{-0.06})$ and $E_{\rm peak} = (508^{+130}_{-84}, 161^{+148}_{-87}, 230^{+46}_{-34})$ for the first, second, and third $\gamma$-ray episodes. The results show mild shallow-to-steep evolution in $\alpha$ and a softening of the second and third emission episodes.  Both model types are consistent with each other and with the values reported by the GBM team \cite{gruber11}; full model parameters are summarized in Table \ref{spectra}. The derived KW fluence in the three emission episodes are $3.05^{+0.24}_{-0.35}$, $1.30^{+0.26}_{-0.14}$, and $8.78^{+0.52}_{-0.40}\times 10^{-5}$\,erg\,cm$^{-2}$ in the 20\,keV to 10\,MeV band.    

Second, we performed a more refined time-resolved analysis on the KW-BAT joint spectra, selecting smaller time bins using a Bayesian Block technique and combining bins as necessary to achieve sufficiently high S/N to properly perform the statistical analysis.  The error bars are larger in the time-resolved analysis and it is difficult to ascertain the true behavior of the spectrum, but the values of both $\alpha$ and $E_{\rm peak}$ are consistent with the time-integrated analysis of the entire first episode and with typical spectral parameters found in other bursts.  Further evidence in support of the softening spectrum is found in the reduction of the hardness ratio between the different KW detectors.  The results are summarized in Table \ref{spectra2}.  

\subsubsection{Interpulse Emission}

We also investigate the interpulse emission and find significant levels of faint emission between the first and second emission episodes ($\sim 7\sigma$ in the KW G1 detector and $\sim 6\sigma$ in the G2 detector).  Since the level of emission is extremely low we cannot perform a time resolved analysis of the entire interval and we take the entire interpulse region as one spectral bin. This emission is best fit by a simple power law with photon index of $-1.73_{-0.12}^{+0.13}$ and over its roughly 530\,s has a fluence of $2.64_{-0.67}^{+1.10} \times 10^{-5}$\,erg\,cm$^{-2}$ in the 20\,keV to 10\,MeV band.  This is a clear indication that the central engine may not cease all activity but may simply suffer from a temporary reduction in accretion rate until later times \cite{lopez10}, further supporting the conclusions from the fractional power density analysis of underlying timescales in the $\gamma$-ray emission. 

\subsection{Energetics}

Using the durations and spectra derived from KW, we calculate the fluence and isotropic energy radiated in $\gamma$-rays.  The total rest frame $1-10^4$ keV isotropic equivalent radiated energy from all three emission episodes is $E_{\gamma, \rm iso} \approx (4.5 \pm 0.09) \times 10^{53}$\,erg, with $0.90^{+0.04}_{-0.03}$, $0.50^{+0.04}_{-0.05}$, and $3.1^{+0.07}_{-0.06} \times 10^{53}$\,erg corresponding to each episode, and we find that all three emission episodes fall within the $2\sigma$ region of the Amati relation \cite{amati02}.  The interpulse segment also emits low-level emission, which when integrated over its long duration radiates an additional $0.74^{+0.07}_{-0.04} \times 10^{53}$\,erg, bringing the total radiated energy up to $5.2^{+0.12}_{-0.09}\times10^{53}$\,erg.

Deriving the kinetic energy contained in the GRB ejecta is a more involved process requiring more detailed broadband afterglow modeling \cite{pandk01} or X-ray afterglow data \citep{fandw01,berger03,lloydronning04,zhang07a}.  Following the theory proposed by Zhang et al. (2007), we determine the spectral regime of the X-ray afterglow and calculate the kinetic energy and radiative efficiency, $E_K$ and $\eta$, where $\eta = E_{\gamma, {\rm radiated}}/E_{\rm total}$.  

A main ingredient in this analysis is the X-ray flux.  Simply extrapolating the observed X-ray light curve back to early times causes the fit to significantly overestimate the amount of flux at early times compared to what is expected from the BAT emission.  We therefore consider two different smoothed broken power law fits which give a more realistic prediction of the level of early-time flux.  The best-fit parameters are $\alpha_1 = 1.06 \pm 0.37$, $\alpha_2 = 2.39 \pm 0.58$, and $t_{\rm break} = 9.57 \times 10^3$\,s with smoothing parameter $s = -3$. This fit still slightly overestimates the expected emission when extrapolating into the BAT band, and we take an approximate $\alpha_1 \approx 0.8 \pm 0.1$ with identical $t_{\rm break}$ and $\alpha_2$ for our best intuitive guess of the trend of the early X-ray afterglow.  We do not claim that this simplistic scenario is the true shape of the light curve but adopt it as a guide for the subsequent efficiency calculation.  

Next, we use the light curve and spectral fits of the X-ray data and the predicted standard model closure relations to ascertain the spectral regime, $\nu_m < \nu_X < \nu_c$ or $\nu_X >$ max$[\nu_m,\nu_c]$.  We use the shallow light curve slope for the early X-ray light curve, $\alpha_1 \approx 0.8 \pm 0.1$, so as to not overestimate the amount of flux and to provide a lower limit on $E_K$.   The slopes are consistent with the constant-density interstellar medium (ISM) model in the $\nu_m < \nu_X < \nu_c$ regime over a large range of $p$ values, or ISM and wind models in the $\nu_X >$ max$[\nu_m,\nu_c]$ regime with $1 < p < 2$.  The very shallow spectral slope of $\beta = 0.49$, however, gives a very low value of $p$ in the latter spectral regime and we therefore assume that the X-ray data lie in the $\nu_m < \nu_X < \nu_c$ regime.  This is consistent with the treatment of shallow $\beta$ by Zhang et al. (2007) and implies $p = 1.98^{+0.23}_{ -0.21}$.  This is not the preferred spectral regime to determine $E_K$ as it is dependent more heavily on the value of $\epsilon_B$.  $E_K$ in units of $10^{52}$\,erg is expressed by Zhang et al. (2007, their Equation 13) for this spectral regime as
\begin{eqnarray}
E_{K,52} & = & \Bigg[ \frac{\nu F_\nu (\nu=10^{18} \rm~Hz)}{6.5 \times 10^{-13}\,{\rm erg}\,{\rm cm}^{-2}\,{\rm s}^{-1}} \Bigg]^{4/(p+3)}
\nonumber \\ 
& \times & D_{28}^{8/(p+3)} (1+z)^{-1} t_d^{3(p-1)/(p+3)}
\nonumber \\ 
& \times & f_p^{-4/(p+3)} \epsilon_{B,-2}^{-(p+1)/(p+3)} \epsilon_{e,-1}^{4(1-p)/(p+3)}
\nonumber \\ 
& \times & n^{-2/(p+3)} \nu_{18}^{2(p-3)/(p+3)},
\label{EK}
\end{eqnarray}
where $\nu F_\nu$ is the flux in the X-ray band, $D_{28}$ is the luminosity distance in units of $10^{28}$\,cm, $n$ is the ambient density, $\nu_{18}$ the observed band in units of $10^{18}$\,Hz, and $f_p$ is a function of $p$ defined as (Equation 10 of Zhang et al. 2007)
\begin{equation}
f_p=6.73 \Bigg( \frac{p-2}{p-1}\Bigg)^{(p-1)} (3.3\times10^{-6})^{(p-2.3)/2}.
\end{equation}

The radiative efficiency of a burst is assumed to be a constant value but the choice of time for calculating the efficiency, $t_d$, varies among bursts. We take $t_d$ to be the time of the second optical peak, $\sim 0.03$\,d, as we are exploring the possibility that this is either approximately the deceleration time or the end of the energy injection from the central engine.  In both cases we want to make sure that there is no significant further addition of energy that will skew the results of the calculation.  We use Equation \ref{EK} with $\epsilon_B = 0.001$, $\epsilon_e=0.1$, $n=1$, $p = 2.1$, and $\nu F_\nu= 4 \times 10^{-10}$\,erg\,cm$^{-2}$\,s$^{-1}$ at the assumed $t_d$ of $\sim 0.03$\,d, and we find a \textit{conservative lower-limit} of $\sim 2 \times 10^{53}$\,erg for $E_K$, which implies $\eta \approx 0.4$, or a relatively inefficient radiator.  The radiative efficiency is also dependent on the value of $p$ and changes dramatically due to the functional form of $f_p$.  The value of $p=2.1$ roughly maximizes Equation 2 and changes to this value will results only in increases of the KE.  For the case where $p \approx 3$, which becomes important for the late-time afterglow modeling, the efficiency drops to the order of $\eta \approx 0.2$.  Decreasing $\epsilon_B$ or $n$ increases the amount of kinetic energy, further lowering the value of the radiative efficiency. Zhang et al. (2007) find that for other bursts in this spectral regime, $\epsilon_B$ is generally very low ($<10^{-4}$) to satisfy $\nu_c > \nu_X$, and could further justify an increase in the estimated kinetic energy of GRB~091024A.  The value of the efficiency is only an estimate, but within the relatively broad range of parameters for this burst we have established the presence of a significant amount of kinetic energy available to power the observed structure of the optical afterglow.

\section{Origin of the Optical Emission}

GRBs for which early optical emission is observed show a range of properties. Some exhibit clear temporal coincidence between optical and $\gamma$-ray features, suggesting a prompt origin \citep{vestrand05,vestrand06,racusin08,guidorzi11a,kopac13}. Others show single peaks or power law decays consistent with the onset or continuation of the afterglow (e.g., Akerlof et al. 1999; Molinari et al. 2007;  Page et al. 2007; Melandri et al. 2008). GRB~091024A has multiple peaks in the $\gamma$-ray and optical bands (Figure \ref{optical_fitting}).

We use a number of cross-correlation tests to see if the optical emission shows any temporal correlation with the $\gamma$-rays. For every step of the KW light curve (2.944 s), we shifted the KW curves along a range of temporal intervals, from $-2000$ to +2000\,s. After shifting, we rebinned the KW curves so as to match the optical binning as closely as possible, and then calculated the Pearson, Spearman, and Kendall correlation coefficients of the optical flux versus $\gamma$-ray rates.  For 1401 trial lags the coefficients are 0.71 (lag $=-490$\,s), 0.78 (lag $=-230$\,s), and 0.62 (lag $=-230$\,s), respectively, where the lag corresponds to the temporal shift of the KW light curve with respect to the optical light curve.  The associated probabilities for these coefficients are of the range of $10^{-3}-10^{-4}$ and correspond to the alignment of the second and third $\gamma$-ray peak with the first optical peak (see Figure \ref{optical_gamma}).  We also stretched the $\gamma$-ray light curve by a scale factor and performed a similar cross-correlation.  We conclude that, despite the richness of temporal structure and significant overlap in the $\gamma$-ray and optical light curves, there is no correlation (or anticorrelation) between the prompt $\gamma$-ray emission and the observed optical peaks. This implies that the optical emission is from a distinct physical process from the prompt $\gamma$-rays and likely consistent with an external-shock origin, as detailed below.  

Further justification for the external-origin hypothesis comes from the morphology of the first optical peak. A diagnostic of possible internal optical emission is the pulse-width$/t_{\rm peak}$ ratio (e.g., Kopa$\check{\rm c}$ et al. 2013), with GRB 091024A showing a value $>$ 1, larger than the typical internal ratio of $<1$.  Alternatively, if the optical peaks are due solely to forward-shock emission, they show consistency with the empirically fit rise and decay slopes of other bursts ($\alpha_{\rm rise} \approx -0.3$ to $-4$ and $\alpha_{\rm decay} \approx 0.6$--1.8) and the empirical $L - t_p$ and $L - E_{\gamma, {\rm iso}}$ relations presented by Liang et al. (2013).

\subsection{Afterglow Modeling}

Having excluded a prompt origin for the optical emission, we now fit the optical light curve with several external synchrotron shock models to determine the nature of the optical peaks, which we conclude are consistent with the presence of strong reverse-shock and forward-shock components as well as late-time rebrightening. We show that the emission is consistent with an intermediate shell thickness regime which can explain the shallow rising light curve at early times, which is a natural consequence of the central engine not ceasing after the first $\gamma$-ray emission episode but continuing to emit at a low level.

\subsubsection{Forward Shock and Refreshed Shock}

Visual inspection of the optical afterglow shows that the first optical peak occurs \textit{before} the second and third $\gamma$-ray emission episodes (Figure \ref{optical_gamma}).  It would be reasonable, then, to expect that the first optical peak evolves like a single GRB with the characteristics of the first emission episode without yet being affected by later emission.  To this end, we model the early optical afterglow with the parameters of the first $\gamma$-ray episode: $E_{\rm iso} \approx 1 \times 10^{53}$\,erg, duration (from KW) $T = 88$\,s.  In this scenario, the first peak is caused by the decelerating fireball from the first emission episode while the second peak is a rebrightening feature caused by the kinetic energy injected into the blastwave from the second and third $\gamma$-ray episodes.  Depending on the burst parameters, specifically the unknown values of the microphysical parameters, the typical frequency of the forward shock (FS), $\nu_{m,f}$, will fall at or around the optical band, $\nu_O$.  In this simple model, if $\nu_{m,f}$ falls at or below the observed band the onset of the deceleration of the fireball and the reverse shock (RS) crossing time will occur simultaneously.

\begin{figure}
\includegraphics[angle=0,scale=0.95]{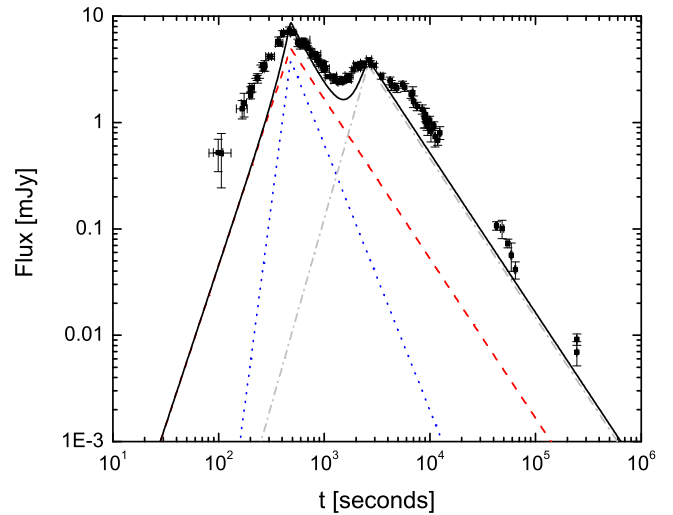}
\caption{Afterglow modeling of GRB 091024A as a forward-shock peak (red; dashed) with simultaneous reverse shock (blue, dotted) followed by a rebrightening peak (gray; dash-dot).  The sum of the components is in black (solid).  $F_{\nu,{\rm max},f} = 5$ mJy, $t_{p,f} = t_{p,r} = 480$\,s,  $F_{\nu,{\rm max}}$ rebrightening $= 3.5$ mJy, $t_{p,\rm rebrightening} = 2600$\,s, $p=3$.} 
\label{FSRB}
\end{figure}

In Figure \ref{FSRB} we show the standard evolution of the FS with a maximal flux from the optical light curve ($t^{3}$ rise, $t^{-3(p-1)/4}$ decay) peaking at 5.5\, mJy combined with the expected RS emission and a rebrightening feature.  We have modeled the latter as a broken power law peaking at 3.5\,mJy, rising as $t^{3.5}$, and decaying with the same slope as the FS, $t^{-3(p-1)/4}$.  The observed late-time slope of $\sim 1.5$ implies $p \approx 3$ and we assume a constant ISM density of 1 proton cm$^{-3}$. Reducing $p$ (e.g., $p \approx 2.2$--2.6) increases the amount of RS flux and makes the slope of the FS and of the rebrightening feature more shallow.  Reducing the peak flux of the FS with this lower $p$ causes the first peak to decrease but also lowers the contribution after the peak, causing the model to further underpredict between the two optical peaks.  Additionally, the shallower slope is inconsistent with the late-time optical data, requiring a mechanism to reduce the late-time flux (e.g., a jet break).

As discussed in \S 3.3, the fireball is likely an inefficient radiator, leaving an ample supply of kinetic energy from the central-engine activity that powered the later $\gamma$-ray emission episodes to refresh the blastwave and cause the rebrightening feature.  The ratio of radiated energy from the combined second and third emission episodes to the first peak is about 3.5, which should approximate the ratio of KE deposited under the assumption that the KE is proportional to $E_{\gamma,{\rm iso}}$.  The ratio of flux at and after the rebrightening peak in Figure \ref{FSRB} is about a factor of $\sim$ 8 above the blastwave solution, which is above the simplistic estimate of the amount of energy injected by the later central-engine activity.  This model has difficulty in explaining the late-time structure of the light curve (i.e., the small rebrightening at $T \approx 5000$\,s), especially if a significant amount of KE is needed to explain the flux of the two large blastwave features.

The early slope of the first peak is not well modeled in either case.  Invoking an intermediate ambient medium density profile can reduce the slope, but there is no clear physical interpretation for such a density profile; furthermore, it affects the afterglow evolution after the crossing of $\nu_m$ \cite{liang12}.  This evidence, in conjunction with the shortfalls discussed above, leads us to consider our next case, that the first optical peak is produced from a reverse shock and the second by the forward shock.

\subsubsection{Reverse Shock and Forward Shock}

In this interpretation, we assume that the first optical peak is caused by the external/reverse shock and the second is caused by the forward-shock emission as the typical frequency of the forward shock, $\nu_{m,f}$, crosses the observed band, $\nu_R$.  A prominent reverse shock has been theoretically predicted \citep{akerlof99,sari99,meszaros99} and seen in a handful of GRBs \citep{kulkarni99,kobayashi00,gomboc08,gendre10,laskar13}.  Using the same parameters for the first $\gamma$-ray episode as above, we model the afterglow with the theoretical predictions of these two components.  Due to the well-sampled optical light curve, we have both the observed times and maximal fluxes of both shock components, which add important constraints that are often missing in this type of analysis. Similarly to the previous section, the FS evolves as $t^3$ at very early times, $t^{1/2}$ as the typical FS frequency approaches the optical band, and then decays as $t^{-3(p-1)/4}$ as the fireball decelerates.  

Depending on the value of the critical Lorentz factor, $\gamma_{\rm crit} = (3E/32\pi n m_p c^2 \Delta_0^3)^{1/8}$, the reverse shock can (thick-shell case, $\gamma_0 > \gamma_{\rm crit}$) or cannot (thin-shell case, $\gamma_0 < \gamma_{\rm crit}$) effectively decelerate the unshocked shell material \citep{kobayashi00,kobayashi03,zhang03,zhang05}.  Here, $\gamma_0$ is the initial Lorentz factor of the shell, $m_p$ is the proton mass, and $\Delta_0$ is the initial shell width, which in the internal-shock model is approximated by the intrinsic duration of the GRB, $\sim cT/(1+z)$ \cite{kobayashi97}.  These cases are the extreme scenarios of relativistic or Newtonian reverse shock, and they provide clear predictions for the rising slope of the reverse shock: $t^{0.5}$ for thick-shell (relativistic) and $t^{3p-3/2} \approx t^{5-6}$ for thin-shell (Newtonian).  

In the case of GRB 091024A, we see an intermediate rising slope of $\sim 2.3$, which clearly deviates from the predictions.  Nakar \& Piran (2004) showed that if a burst falls into an intermediate regime between fully relativistic and Newtonian, the rising slope of the reverse shock can change dramatically, characterized by the dimensionless parameter $\xi = (l/\Delta_0)^{1/2}\gamma_0^{-4/3}$, where $l = [3E/(4\pi n m_p c^2)]^{1/3}$, $\Delta_0$ is the shell width, and $\gamma_0$ is the initial Lorentz factor.  The cases $\xi << 1$ and $>> 1$ correspond to relativistic and Newtonian, respectively.  An intermediate-regime fireball would require longer-lasting central-engine activity, which is justified in the case of GRB 091024A since we see low-level emission between the first two $\gamma$-ray episodes (see \S 3.2.2). 

Having observed the rise of the reverse shock ($\alpha_{rise} \approx -2.3$), we can estimate the value of $\xi$ from the numerical results of Nakar \& Piran (2004):
\begin{equation}
\alpha_{\rm rise} \approx N_{\alpha} \bigg[ 0.5 + \frac{p}{2} (\xi - 0.07 \xi^2) \bigg],
\end{equation}
where $N_\alpha$ is a numerical constant $= 1.2$. Depending on the value of $p$, $\xi$ can range from $\sim 1$ to 1.4; it is consistent with the intermediate-regime treatment where $\xi \approx 1$.

In a complementary analysis, Harrison \& Kobayashi (2013, hereafter HK13) used hydrodynamical simulations to study afterglows with significant external/reverse-shock emission and found that previous treatments significantly underestimate the amount of magnetization in the RS by as much as two orders of magnitude, especially in intermediate regions where $\xi \approx 1$. Having larger magnetization in the RS significantly lowers the initial Lorentz factor, helping to move the RS from a highly relativistic (thick-shell) regime that significantly decreases the reverse-shock slope before the RS crossing time. The inclusion of magnetization is further rationalized by the discovery of 10\% polarization in the early afterglow of GRB 090102 \cite{steele09}, proving the existence of large-scale ordered magnetic fields in that, and likely other, bursts.  HK13 provide updated magnetization expressions based on their hydrodynamical simulations with numerical corrections to the theoretical framework of Zhang, Kobayashi, \& M\'{e}sz\'{a}ros (2003) and Zhang \& Kobayashi (2005), and to the treatment of GRB 061126 by Gomboc et al. (2008).  

Since the duration of the central-engine activity is longer than simply the observed initial $\gamma$-ray episode, we can estimate the duration of the central engine needed to create the observed shallow slope from
\begin{equation}
T \approx t_{p,r} \bigg(\frac{\xi^2}{5} + 1 \bigg)^{-1},
\end{equation}
where $t_{p,r}$ is the peak time of the reverse shock and $T$ is the duration of the emission.  Another consequence of the intermediate regime is the estimation of the initial Lorentz factor, $\gamma_0$.  In the thin-shell regime, we can estimate $\gamma_0$ from the crossing time of the shell, $t_x$.  We observe this peak at $480 \pm 19$\,s from the onset of the GRB, and theoretically it should occur at $t_x \approx (\gamma_x/ \gamma_{\rm crit})^{-8/3}T$, where $\gamma_x$ is the Lorentz factor at the shock-crossing time, $\sim$ min$[\gamma_0,\gamma_{\rm crit}]$ \citep{sari99,kobayashi00,zhang03}.  The correction factor to the deceleration time in the thin-shell regime from HK13 is simply $C_t = 2^{-4/3} x^{-8/3}$, where $x = \gamma_d/\gamma_0$, the ratio of the Lorentz factor of the shocked shell material relative to the  unshocked shell and the initial Lorentz factor. The value of $x$ in turn depends on $\xi$ as
\begin{equation}
\xi^2 \approx \frac{ 24x^{8/3} }{2^{2/3} (1-x^2)(2+3x+2x^2)}.
\end{equation}

Next, we can estimate the magnetization parameter
\begin{equation}
R_B = \bigg( \frac{R_F^3 \gamma_0^{4\alpha-7}}{C_F^3 C_m^{2(\alpha-1)}R_t^{3(\alpha-1)}} \bigg)^{2/(2\alpha+1)},
\end{equation}
\noindent
the typical frequency of the RS
\begin{equation}
\nu_{m,r}(t_x) \approx C_m \gamma_0^{-2} R_B^{1/2} \nu_{m,f}(t_x), 
\end{equation}
\noindent
and the maximal flux of the reverse shock at the peak time and at the typical frequency
\begin{equation}
F_{\nu,{\rm max},r}(t_x) \approx C_F \gamma_0 R_B^{1/2} F_{\nu,{\rm max},f}
\end{equation}
\noindent
(Kobayashi \& Zhang 2003; Zhang et al. 2003; Zhang \& Kobayashi 2005; HK13), where $C_F$ and $C_m$ are the numerical correction factors from HK13, $\alpha = (3p+1)/4$ is the decay slope of the RS, $R_t = t_{p,{\rm FS}}/t_{p,{\rm RS}}$ is the ratio of the FS peak time to RS peak time, and $R_F = F_{\nu,{\rm max,RS}}/F_{\nu,{\rm max,FS}}$ is the ratio of the RS peak flux and FS peak flux (Gomboc et al. 2008; HK13). The values of $R_t$ and $R_F$ can be measured directly from the optical light curve.

\begin{figure}
\includegraphics[angle=0,scale=0.95]{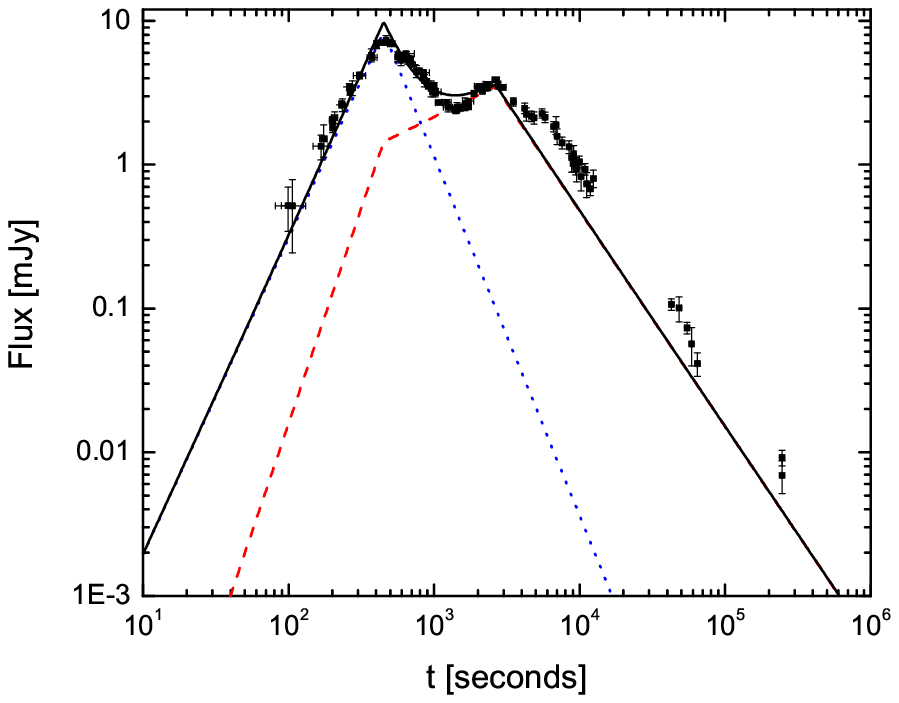}
\includegraphics[angle=0,scale=0.95]{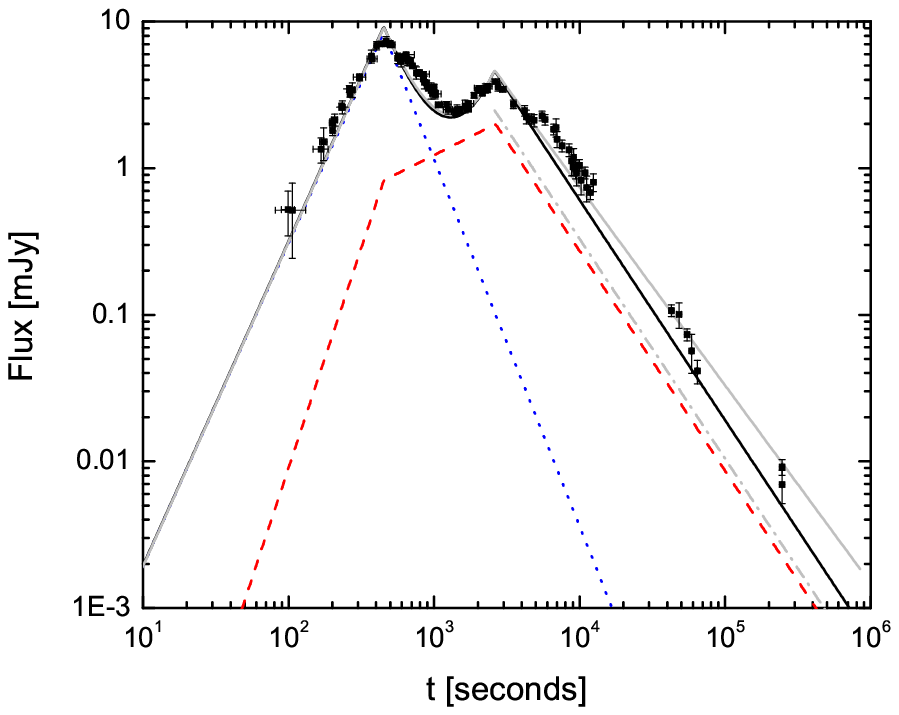}
\caption{Afterglow modeling of the optical light curveof GRB 091024A assuming that the first optical peak is due to emission from a magnetized reverse shock  (blue; dotted) in an intermediate regime between the thin- and thick-shell approximations, and that the second peak is either the forward shock (red; dashed) or forward shock and simultaneous rebrightening (gray; dash-dot).  The sum of all components is given in black (solid).  (a) $F_{\nu,{\rm max},f} = 3.5$ mJy, $p=3$ (b) $F_{\nu,{\rm max},f} = 2$ mJy and $F_{\nu,{\rm max, rebrightening}}= 2.5$ mJy, $p=3$ (solid, black) and $p=2.8$ (solid, gray)}
\label{RS}
\end{figure}

Figure \ref{RS} shows two examples of model fits to the optical light curve of GRB 091024A.  Using the RS rise slope we calculate $\xi \approx 0.95$ for $p=3$.  The first panel shows the model predictions using the observed parameters $t_x = t_{m,r} = 480$\,s, $F_{\nu,m,r} \approx 7.5~\rm mJy$,  $F_{\nu,m,f} \approx 3.5$\,mJy, $t_{p,f} \approx 0.03$\,days, and $n_0 = 1$\,proton cm$^{-3}$. This implies $\epsilon_e \approx 0.18$, $\gamma \approx 115$, $R_B \approx 100$, and $T \approx 380$\,s.  In addition, the rising and decay slopes of the RS better approximate the observed light curve.  This model, however, washes out some of the structure between the two optical peaks, due to the $t^{1/2}$ component of the FS emission before the peak, and is problematic from an energetics point of view since the modeling assumes that the driving force of the light curve is only the first $\gamma$-ray emission episode.  If a significant fraction of the kinetic energy emitted by the burst is yet to be deposited into the blastwave, this solution does not provide the flexibility for energy injection into the blastwave, apart from the small feature at $\sim 5000$\,s.

In order to address this problem, we add a rebrightening component near the time of the second optical peak.  In this case, the second optical peak will be a superposition of the two components, allowing us to decrease the importance of the FS flux and see the effect on the modeled light curve (Figure \ref{RS}, panel 2).  Lowering the peak flux of the FS increases the magnetization parameter but only marginally affects the RS peak flux.  Similar to the FS analysis, we introduce a rebrightening component that evolves as $t^{3.5}$ and $t^{-3(p+1)/4}$ and peaks near the forward-shock peak time of 2600\,s.  We model the RS and FS as discussed above, with the exception of varying the peak flux of the FS.  The second panel of Figure \ref{RS} shows a model fit with the the same $\xi$ and $p$, a FS peaking at 2 mJy and the rebrightening feature at 2.5 mJy, implying $\epsilon_e \approx 0.18$, $\gamma \approx 115$, $R_B \approx 185$, and $T \approx 380$\,s.  This also adds flexibility for the energy injection in the form of the rebrightening peak, which is comparable to the FS emission, still leaving room for further injection of energy to create the feature at 5000\,s.  The ratio of flux from the total solution to the FS solution is about a factor of 2.3.  Decreasing the value of $p$ slightly shallows all the decay slopes, but $p$ less than $\sim 2.8$ begins to exceed the flux of the late-time data points.  With a $\xi = 1.03$, this implies $\epsilon_e \approx 0.21$, and it has the additional effect of decreasing the magnetization and duration to $\sim 130$ and 370\,s, respectively.

The behavior of the forward-shock evolution at very early times (before the peak time) in this intermediate model is not well understood.  If it behaves similarly to the RS in that the slope lies somewhere between the extreme solutions of the thin- and thick-shell cases, it is possible that the FS slope is shallower than the expected $t^3$ evolution in the thin-shell regime.  Decreasing the FS slope to $\sim 1$ has only a small effect on the \textit{total} flux of the light curve, and the behavior lies within the error bars of the observed data points.  

\section{Ultra-long GRB Sample}

GRB 091024A is one of a handful of bursts with interesting and very long-duration prompt $\gamma$-rays.  Its prompt emission has strong (i.e., high flux) $\gamma$-ray emission totaling $\sim 700$\,s that is interrupted by long periods of low, almost background-level emission.  In addition to its prompt $\gamma$-ray emission, this burst has unique and well-sampled optical emission showing multiple peaks and bumps that we have interpreted as emission from a magnetized external/reverse shock followed by a forward shock with significant energy injection.  

Next, we review other ultra-long GRBs with $\gamma$-ray duration $\gtrsim 1000$\,s discussed in the literature and present them grouped by similar overarching themes and characteristics \textit{in their $\gamma$-ray emission}.  All of these bursts show $\gamma$-ray emission at times beyond what is expected for most GRBs ($\gtrsim$ 1000\,s), whose distribution of $t_{90}$ durations peaks at a few tens of seconds and extends to a few hundred seconds, depending on the energy range of the detector \citep{virgili12,qin13}.   A summary of relevant observations is found in Table \ref{sample}.  Although a number of possible very long-duration events were discovered with CGRO/BATSE (Tikhomirova \& Stern 2005), we do not include these due to the inability to robustly associate the many $\gamma$-ray episodes with one GRB \cite{palshin12}.

\subsection{Interrupted Emission}

GRBs within this category have episodes of prompt $\gamma$-ray emission separated by long stretches of low-level or quiescent activity lasting hundreds of seconds, that require vast reductions in central-engine activity.  Bursts that show this behavior are: GRB 080407A, GRB 091024A, GRB 110709B (see Appendix for full details).  Additionally, most of the BATSE bursts presented by Tikhomirova et al. (2005), assuming the emission episodes could be linked to the same GRB, have this type of behavior (durations of 500 to 1300\,s). 

Even with a smaller sample size than the other category of bursts, there is still a range of afterglow properties.  GRB 091024A and GRB 110709B both show very high-flux $\gamma$-ray emission and normal to weak X-ray afterglows.  We have previously discussed the optical properties of GRB 091024A, and GRB 110709B is classified as a dark burst. 

In this case, ``duration'' has less significance than in the continual emission case.  Even if the central engine has not completely ceased to emit, as we have shown in the spectral analysis of GRB 091024A, the large interpulse interval could change, for example, the evolution of the expected afterglow emission. This is exemplified in GRB 091024A, whose intermediate thin/thick-shell regime modifies the evolution of the reverse-shock emission.  It might also provide insight into the distribution of matter in the accretion disk or how matter is fed into the central engine (e.g., a fall-back accretion model; Wu et al. 2013 and references therin).  Long quiescent times have also been interpreted in the literature as outliers to the log-normal distribution of interpulse times, possibly indicating a different mechanism than other intervals \citep{quilligan02,nakar02,drago07}.  There is also weak evidence for a correlation between the pulse width and the following interval \cite{nakar02}.  Alternatives include the formation of quark phases \cite{drago08} or changes to the distribution of ejected shells from the central engine (e.g., Ramirez-Ruiz, Merloni, \& Rees, 2001).  GRBs 080407A, 091024A, and 110709B have long interpulse regions lasting $\sim 1400$, 500, and 600\,s, respectively.  The BATSE bursts discussed previously have interpulse episodes of a similar timescale, with a few bursts reaching 600--1400\,s.  All bursts tend to have emission episode durations of 50--200\,s.  An intermediate case is found in bursts like GRB 010619A, discovered by \textit{Beppo}SAX \citep{frontera09,guidorzi11a}.  Its duration is shorter than the rest of the sample ($T_{90} \approx 450$\,s), but it shows many of the characteristics of this class, including intense emission episodes and significant periods of quiescence between emission episodes. This behavior is mimicked in various other bursts, e.g., GRBs 001213A and 121217A.

Previous works have not discussed the implications of selection bias on ultra-long GRBs with interrupted emission.  In this case $t_{90}$ is not as meaningful, since the emission drops to nearly background level between emission episodes.  In the case of GRB 091024A, it could be argued that a 1200\,s ``duration'' is misleading since more than half of that time is near-background emission or quiescence.  Other properties, like the peak flux, are also different in bursts with continuous emission.  Instead of low-level continuous emission peaking on the order of 0.1--0.5\,ph\,cm$^{-2}$\,s$^{-1}$ for \textit{Swift}-detected bursts, these have emission episodes that range from 50\,s to 200\,s and peak fluxes of $\sim 10$--1000\,ph\,cm$^{-2}$\,s$^{-1}$.  When the individual emission episodes are plotted on the fluence--$t_{90}$ and 1-s peak flux--$t_{90}$ planes, they are more consistent with the population of ``normal'' long-duration GRBs, increasing the chances that they will be detected (see Figure \ref{fluence}).  Biases in the detectability of \textit{all} the associated emission episodes now become important.  In the four post-BATSE bursts that show interrupted emission, the first emission episode is either comparable to or brighter than the subsequent peaks.  This is exemplified by GRBs 091024A and 110709B that triggered \fermi-GBM and \textit{Swift}, respectively, on both their early and late emission episodes.  The collection of BATSE bursts show similar behavior although not as pronounced.

\subsection{Continuous Emission}

\begin{figure*}
\includegraphics[angle=0,scale=1.05]{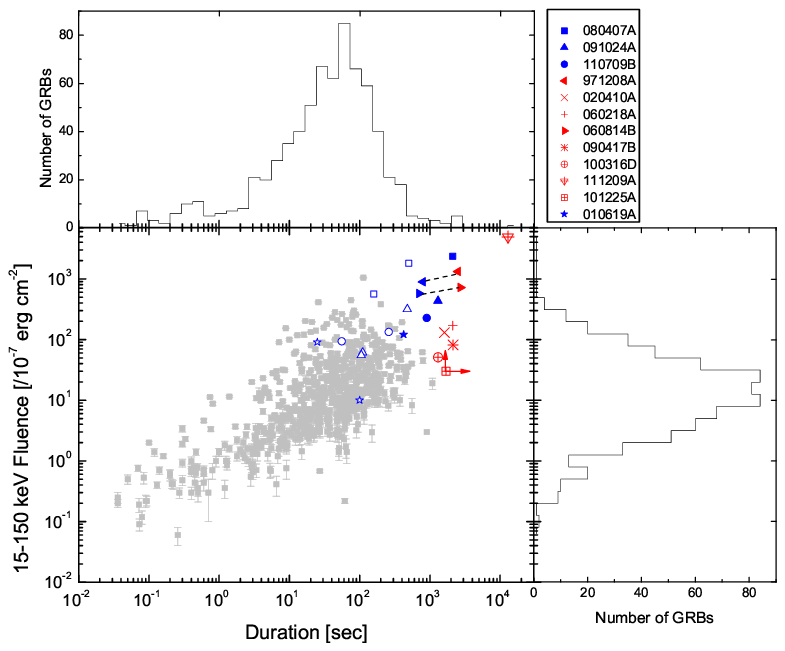}
\includegraphics[angle=0,scale=0.8]{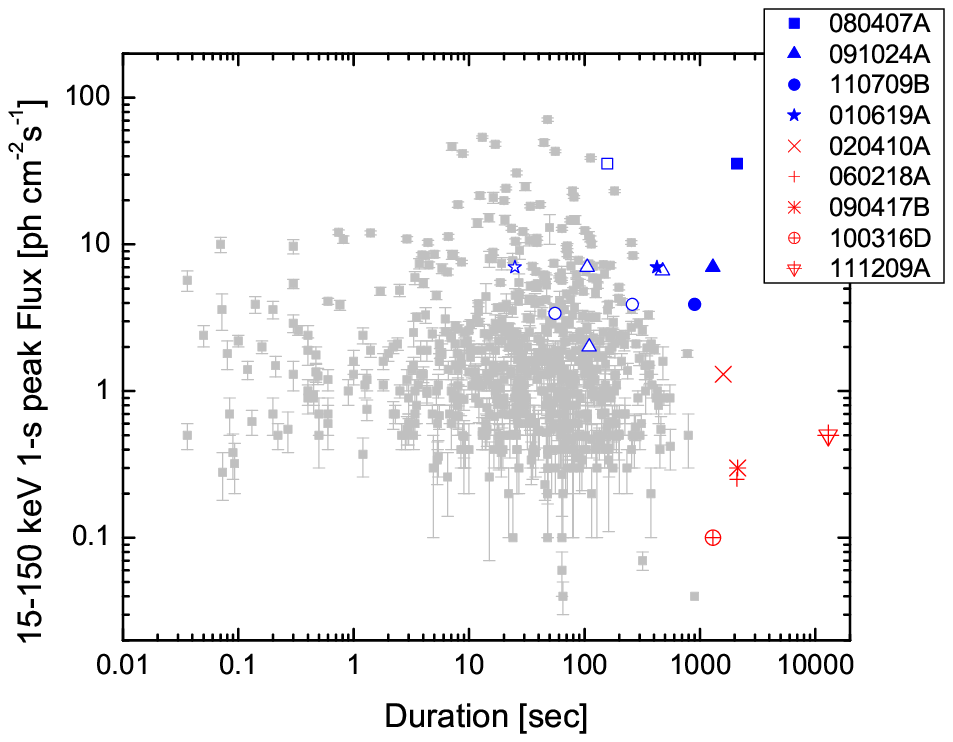}
\caption{Distribution of ultra-long GRBs in the fluence--$t_{90}$ and 1-s peak flux--$t_{90}$ planes.  Light-gray points are all \swift~ detections, red are bursts with continuous $\gamma$-ray emission, and solid blue are bursts with interrupted emission (as discussed in \S5).  Open blue symbols represent the multiple episodes of bursts with interrupted emission as opposed to the duration spanning all then episodes. The dashed lines between single-pulse GRBs 971208A and 060814B indicate the fluence calculated using only the bright peak and the fluence including the extended tail emission.  Values for GRB 101225A are upper limits as the burst was in progress when it entered the \swift\ field of view.  The true duration could be $\gtrsim 10^4$\,s.}
\label{fluence}
\end{figure*}

This subset of ultra-long GRBs shows long-lasting continual emission from the central engine manifested as a single fast-rise exponential decay (FRED)-like pulse or one or more broad overlapping pulses.  These include: GRBs 840304A, 971208A, 060218A, 060814B, 090417B, 100316D, 101225A, and 111209A (see Appendix for full details).  A unifying characteristic of this group is their long-lasting but very low-level emission, as highlighted in Figure \ref{fluence}.  This weak emission is likely the  greatest limitation to the detection of this type of very long event, which inhibits their observation at high redshift and is likely a contributing factor to the large gap in detections between a few 1000\,s and $10^4$\,s duration bursts.

GRBs 971208A, 060218A, 060814B, and 100316D all show the simplest $\gamma$-ray light curve, one broad or FRED-like pulse lasting anywhere from 1300 to 3000\,s.  Slightly more complicated, with a broad $\sim 200$\,s initial peak followed by a 1000\,s tail, is the earliest documented ultra-long GRB discussed in the literature, GRB 840304A.  Morphologically, FRED-like emission and the superposition of peaks together with longs tails is not uncommon.  This behavior was seen by Giblin et al. (2002) in 40 BATSE GRBs and is common in many \textit{Swift}-era bursts. 

GRBs 090417B and 111209A show more complex and variable emission, with multiple overlapping peaks.  The former had continuous emission, particularly in the softer BAT bands, for roughly 2000\,s, while the latter emitted in $\gamma$-rays for nearly 15 ks.  GRB 101225A is a weak burst detected over various \swift~orbits with a redshift of $z = 0.847$ assigned from the spectrum of a dim optical counterpart at the position of the GRB \cite{levan13} and lower limits on the duration ($> 1650$\,s) and fluence \cite{racusin11}.

GRB 020410A  has low-flux $\gamma$-ray emission and significant emission in X-rays, having been discovered by \textit{Beppo}SAX (2--9\,keV) and later in a ground analysis by KW. Its $\gamma$-ray emission has various pulses and evidence of low-level excess in the softest band of KW out to $\sim 2500$\,s. The pulses are temporally similar to bursts with interrupted emission and correspond with pulses observed by \textit{Beppo}SAX, with interpulse periods lasting 50--100\,s.  This near-background interpulse period is likely not an indication of central-engine quiescence but a detectability problem, since this burst has a flux near the threshold of KW and has corresponding structure in the X-ray band.  It is then possible that this characteristic of continued central-engine activity extends to bursts with interrupted emission, like our analysis of GRB 091024A indicates, and highlights the likely overlap of both categories.

In addition to the prompt $\gamma$-rays, about half of these bursts also have multi-wavelength afterglows.  GRBs 971208A and 060814B do not have detected afterglows and apart from their duration, have no indication (e.g., spectra) that they are intrinsically different from other GRBs \citep{giblin02,palshin08}.  GRB 090417B is an optically dark burst with a canonical X-ray light curve \cite{zhang06}, and GRB 111209A is a seemingly extreme version of the former.
The long $\gamma$-ray and X-ray emission of GRB 111209A lasts about 20\,ks and has been interpreted as emission from the collapse of a blue supergiant, while the late-time optical afterglow of GRB 111209A has been interpreted within the context of the external-shock model as emission from the external/forward shock or possibly the external/reverse shock \cite{gendre12,stratta13}.  GRB 101225A also shows long-lived X-ray and optical/infrared emission together with its weak $\gamma$-rays.  

With different behavior from the rest, GRBs 060218A and 100316D are nearby bursts that have long-lived X-ray emission with thermal components\footnote{In a reanalysis of the XRT data, Margutti et al. (2013b) show that the significance of this thermal component is substantially reduced.} and very low $E_{\rm peak}$ values \citep{liang06a,starling11}.  Various theoretical interpretations have been invoked to explain X-ray flashes and soft events, including off-axis viewing angle \citep{liang06a,guidorzi09}, shock breakout emission \citep{liang06a,nakar12,bromberg11}, and sub-energetic explosions (Soderberg et al. 2006, Margutti et al. 2013b). Regardless of the interpretation, the observations seem to imply that events similar to these GRBs are set apart in their prompt and afterglow properties and there is mounting evidence that they may require a different description within the framework of classical GRBs.

\section{Discussion}

We have presented a multi-faceted analysis GRB 091024A as an example of an ultra-long GRB, and in the context of the sample of other such objects in order to shed light onto the population as a whole.  This burst falls into a general category of bursts that show multiple episodes of $\gamma$-rays separated by large periods of quiescence or low-level emission.  By performing a power-spectrum analysis we have shown that these episodes exhibit similar variability timescales, firmly tying them to similar central-engine activity.  Spectrally, the episodes do not show extraordinary activity compared to other GRBs, the KW data revealing $E_{\rm peak}$ values $\sim 300$\,keV, mild spectral evolution, and lower hardness ratio in later emission episodes.  

The well-sampled early optical afterglow displays three bright peaks beginning just after the first $\gamma$-ray episode.  These peaks are not coincident and show no correlation with the $\gamma$-ray peaks, nor do they show tracking behavior like the ``naked-eye burst'' GRB 080319B \cite{racusin08}, which would imply an internal origin.  Since the first optical peak begins before the final $\gamma$-ray episodes it is natural to link it to the first episode.  Our detailed afterglow modeling shows that this burst can be well modeled by a peak caused by a highly magnetized reverse shock, followed by the onset of the forward shock with added energy injection.  Since GRB 091024A has similar spectral characteristics as other GRBs with shorter interpulse emission and has an afterglow that can be reasonably well explained within the framework of a magnetized reverse shock, we argue that this GRB is not fundamentally different but likely an event in the tail of the distribution of ``classical'' long GRBs.  Whether or not this and other bursts with very long gamma-ray duration are a separate population is a relatively new topic recently brought to light by discussions on the observations of GRB 111209A \cite{gendre12,stratta13} and 101225A \cite{levan13}.  \subsection{Statistical test}

We now examine the probabilities of discovering the number of observed events within the assumption of a log-normal distribution of durations to test whether the current observations require the definition of a distinct new class of GRBs. For reference, we show the 1-D duration, 1-D fluence, and 2-D fluence-duration and flux-duration distributions in Figure \ref{fluence}.  We selected the subset of the 591 GRBs \textit{detected by Swift} with log(duration) $> 0.7$ ($\gtrsim 5$\,s) to avoid contamination from short bursts and fit this with a normal distribution (mean = 1.67, $\sigma$ = 0.51).  We then performed a $\chi^2$ test with the expected normal distribution, yielding a $\chi^2$ of 20.9 for 11 degrees of freedom, which corresponds to a probability of accepting the null hypothesis of normality to $3.4\%$.  Truncation of the normal distribution at short durations may contribute to the deviation from normality, and performing this analysis with a truncated normal distribution having similar parameters yields slightly higher probabilities ($\chi^2=19.4$, $P = 5.4$\%).  The largest deviations from the expected distribution as probed by the $\chi^2$ test, however, are not from the bin containing the longest GRBs.  The expected number of GRBs with duration longer than $\sim 630$\,s is approximately 9.75 events, similar to the 11 \swift-observed events.  Reducing the total to 9 bursts, simulating the absence of GRB 060218A and 100316D from the sample, does not significantly change the fit or probabilities.  Only increasing our sample of observed bursts will allow us to fully probe this end of the distribution with a higher resolution. Since we are not able to robustly reject the simple null hypothesis of normality, this test is suggestive that, from the properties of the duration distribution, the observed events are sampled from the tail of the distribution and that there is no current justification for the more complex hypothesis of a separate population (i.e., Occam's razor).  

\subsection{Progenitor models} 

Large-mass progenitors have been recently invoked to produce a variety of interesting long-lived structures in $\gamma$-ray and X-ray-rich bursts (e.g. Peng et al. 2013, Nakauchi et al. 2013), but the relationship between the amount of material present for accretion, the unknown mechanism to start and stop the accretion flow, and how these scale with stellar mass and composition of the progenitor star are not well constrained.  Lazzati et al. (2010, 2012) also caution on the difficulty of associating a central engine activity with the $T_{90}$ duration of a given GRB and the possible effects of viewing angle on the ultimate duration of the burst.  As an example of an alternative model to a significantly larger progenitor star, Wu et al. (2013) propose a fallback accretion model interpretation (e.g. Kumar et al. 2008a, 2008b) for this type of long-lived central engine activity. Observationally, X-ray rich bursts that imply long-lived central engine activity appear to exhibit many similar qualities as `normal' GRBs \cite{feroci01,nicastro04,intZand03}.

In a recent review, Woosley (2011) and Woosley \& Heger (2012) have theorized that the outer layers of a star may have sufficient angular momentum to form a disk, which would appear similar to a GRB jet, but longer and fainter provided the mass-loss rate is not high.  They predict durations of $10^4-10^5$\,s for both supergiant star and Wolf-Rayet binary progenitors, attributing GRB 101225 and similar bursts (i.e. 111209A) to one of these scenarios.  Other studies examine the possibility that at least some GRBs are produced from binary pairs (e.g. Podsiadlowski 2007; Podsiadlowski et al. 2010), providing a viable alternative scenario to the blue supergiant hypothesis.  The requirement that the jet fully penetrate the envelope of the star has proven challenging to models with large progenitors \cite{matzner03}, but recent simulations indicate that  this may be possible under certain conditions (e.g. Suwa \& Ioka 2011; Nakauchi et al. 2012, 2013).  Conversely, Nakauchi et al. (2013) also find that for different values of the (highly uncertain) jet efficiency, wider weaker jets ($>$ 18-24 degrees) have difficulty breaking out of the stellar envelope for several of their blue supergiant models, adding further constraints to this progenitor model.  This becomes particularly important for bursts such as GRB 111209A whose derived jet opening angle is about 23 degrees \cite{stratta13}.  
\section{Conclusions}

The broadband data on GRB 091024A have provided a rich and detailed test of the underlying central engine that powers these interesting and energetic phenomena.  In summary:

\begin{itemize}
\item{GRB 091024A has $\gamma$-ray emission lasting for about 1200\,s that is separated into three separate emission episodes, with weak interpulse emission detected between the first and second episodes.}
\item{We report the spectroscopic redshift of this burst as $z=1.0924 \pm 0.0002$.}
\item{The rich optical data set has three peaks that are not coincident in time with the gamma-ray emission.  We interpret these peaks as emission from a highly magnetized external/reverse shock  in an intermediate shell-thickness regime followed by the forward shock peak and rebrightening feature.}
\item{GRB 091024 shares many properties with `classical' GRBs and is likely an event in the tail of the distribution of long GRBs.}
\item{We analyze a sample of ultra-long GRBs discovered to date. These show natural variety in their characteristics and can be broadly described in two categories: bursts with interrupted and continuous $\gamma$-ray emission. With the likely exception of GRBs 060218A and 100316D, the properties of this sample do not yet provide a strong statistical motivation for defining a new, distinct population of GRBs and instead suggest ultra-long GRBs represent the tail of the duration distribution of the long GRB population.}
\end{itemize}
Further diversified observations at many wavelengths and timescales are essential to further understand the nature of these enigmatic events. 
\\

\acknowledgments

F.J.V. acknowledges support from the UK Science and Technology Facilities Council.  C.G.M. acknowledges funding from the Royal Society, the Wolfson Foundation, and the UK Science and Technology Facilities Council. We are grateful for excellent staff assistance at the various observatories where we obtained data. The Liverpool Telescope is operated by Liverpool John Moores University at the Observatorio del Roque de los Muchachos of the Instituto de Astrof\'{i}sica de Canarias.  The Faulkes Telescopes, now owned by the Las Cumbres Observatory Global Telescope network, are operated with support from the Dill Faulkes Educational Trust. KAIT and its ongoing operation were made possible by donations from Sun Microsystems, Inc., the Hewlett-Packard Company, AutoScope Corporation, Lick Observatory, the NSF, the University of California, the Sylvia and Jim Katzman Foundation, and the TABASGO Foundation. Some of the data presented herein were obtained at the W. M. Keck Observatory, which is operated as a scientific partnership among the California Institute of Technology, the University of California, and NASA; the Observatory was made possible by the generous financial support of the W. M. Keck Foundation. 
The Konus-Wind experiment is supported by a Russian Space Agency contract and RFBR grant 12-02-00032-a. 
Swift, launched in November 2004, is a NASA mission in partnership with the Italian
Space Agency and the UK Space Agency.
A.V.F.`s group at UC Berkeley has received generous financial assistance from Gary and Cynthia Bengier, the Christopher R. Redlich Fund, the Richard and Rhoda Goldman Fund, the TABASGO Foundation, NSF grant AST-1211916, and NASA/{\it Swift} grants NNX10AI21G and NNX12AD73G.
This work made use of data supplied by the UK {\it Swift} Science Data Centre at the University of Leicester.

\begin{appendix}

\section{Ultra-long gamma-ray bursts}

Below we detail the observations of the ultra-long GRBs with $\gamma$-ray durations longer than $\sim 1000$\,s discussed in the literature.  In parentheses we denote how we classified each burst, either as having continuous or interrupted $\gamma$-ray emission.  Some of this information is summarized in Table \ref{sample}.

\subsection{GRB 840304A (Continuous)}

GRB 840304A is the earliest documented ultra-long GRB reported in the literature \citep{klebesadel84}.  It was detected by Pioneer Venus Orbiter, International Cometary Explorer, and Vela 5B and its light curve consists of a 2 broad peaks lasting $\sim 200$\,s followed by a 1000\,s tail.  The fluence is $2.8\times 10^{-3} \rm ~erg~cm^{-2}$.  No afterglow is detected.

\subsection{GRB 971208A (Continuous)}

Detected by CGRO/BATSE \cite{connaughton97,giblin02}, KW \cite{palshin08}, and \textit{Beppo}SAX \cite{frontera09,guidorzi11b} , the light curve of this burst is one extremely bright and long-lasting fast-rise exponential decay (FRED)-like pulse lasting 2500\,s  and with a fluence of $2.55 \pm 0.11 \rm ~erg~cm^{-2}$ in the 15-1000 keV band \cite{palshin08}.  This long pulse can be separated into a strong initial pulse that contains a large fraction of the fluence, followed by a long-lasting and low-level tail, similar to GRBs 840304A and 060814B. This behavior was seen by Giblin et al. (2002) in 40 BATSE GRBs and is also common in many \textit{Swift}-era bursts.  The spectrum shows hard to soft spectral evolution as a function of time, again similar to GRB 060418B, and the time-integrated $E_{\rm peak}$ (over the entire burst interval) is a typical $144 \pm 12$ keV.  The spectrum of the intense initial peak (485\,s) is slightly harder with a peak energy of $165 \pm 7$ keV \cite{palshin08}.

\subsection{GRB 020410A (Continuous)}

This burst was detected by \textit{Beppo}SAX in X-rays and as an offline detection by KW \cite{nicastro04} and has four overlapping peaks lasting $>$1300\,s (2-10 keV) and fluence $>4.7\times10^{-6}\rm ~erg~cm^{-2}$.  The duration of the X-rays is only an upper limit as Earth occultation caused the observations to cease, but a reconstruction of the partially observed final peak and coincidence with the KW observations give an estimate of $\sim 1550$\,s for the duration and $2.8 \times 10^{-5}  \rm ~erg~cm^{-2}$ for the 15-1000 keV fluence of this burst, respectively.  A possible 2.5$\sigma$ excess in the softer channels from $T_0+1500$ to $T_0+ 2500$\,s is also reported.   This burst shows clear overlap of structure in the $\gamma$-rays and long X-ray emission, showing it is plausible that the X-ray structure in other long-lived X-ray-rich bursts with shorter $\gamma$-ray durations \cite{intZand03} could in fact be caused by a process similar to that which creates the weak $\gamma$-rays.  No redshift is found for this burst but it is estimated as $0.9 < z < 1.5$ from the Amati relation \cite{amati02}.  Late-time optical emission is detected and can be possibly explained by refreshed shocks or the emergence of a supernova component \cite{levan05}.  By assuming this emission is from a supernova, Levan et al. (2005) estimate the redshift as $z \sim 0.5$. No radio afterglow was detected to $>120 \mu$Jy at 8 GHz.

\subsection{GRB 060218A (Continuous)}

This burst is very similar to the description of GRB 100316D (See \S A.9 for further discussion).  Its emission lasts for approximately 2100\,s in the \swift/BAT band, has an $E_{\rm peak} = 4.9^{+0.4}_{-0.3}$ keV, and shows indications of a thermal component \cite{soderberg06,campana06,liang06a}.  The fluence is $1.7\times10^{-5} \rm ~erg~cm^{-2}$ in the 15-150 keV and corresponds to an $E_{\rm iso}$ of $(6.2 \pm 0.3)\times10^{49}$ erg \cite{campana06,soderberg06} at $z= 0.0331$ \cite{mirabal06}.  The optical light curve is consistent shows two prominent peaks, the latter being consistent with the emergence of a SN \cite{campana06}.  Radio observations are indicative of a mildly relativistic explosion in a quasi-spherical flow, with a jet opening angle of $> 80^\circ$ and Lorentz factor of 2.3 at about 5 days \cite{soderberg06}.

\subsection{GRB 060814B (Continuous)}

Showing similarity to GRB 971208A, the light curve of this burst shows a single fast-rise exponential decay (FRED)-like peak lasting about 2700\,s, with the most intense portion of the peak lasting $\sim$ 700\,s.  This burst was detected in its entirety by KW and the initial portion by Ulysses, Mars Odyssey (HEND), Suzaku-WAM, and INTEGRAL-SPI-ACS \cite{palshin08}.  Continued low-level emission that is likely associated with this burst was also detected in the softest KW channels, lasting on the order of $10^4$\,s.  The spectrum shows hard-to-soft evolution and has typical peak energies, $374 \pm 30$ keV for the initial pulse and $341 \pm 61$ keV for the entire 2700\,s pulse duration, giving a derived fluence of $(2.35 \pm 0.22)\times 10^{-4} \rm ~erg~cm^{-2}$ in the 18-1170 keV range \cite{palshin08}.

\subsection{GRB 080407A (Interrupted)}

This burst is an example of a burst with interrupted emission detected by KW, showing two separate emission episodes lasting  $\sim$ 160\,s and $\sim$ 400\,s, respectively, separated by $\sim$ 1500\,s \cite{palshin12}.  Portions of this burst were detected by a variety of other spacecraft (see Pal'shin et al. 2012 and references therein). The first episode shows some sub-structure, with a hard and bright spike ($E_{\rm peak} = 325^{+29}_{-25}$ keV) followed by a softer peak of similar duration ($E_{\rm peak} = 114^{+77}_{-44}$ keV), and a total derived fluence of $(1.43 \pm 0.04)\times 10^{-4} \rm ~erg~cm^{-2}$ (20-1000 keV).  The second episode, unfortunately, does not have spectral information but has an estimated fluence of $\sim 3 \times 10^{-4} \rm ~erg~cm^{-2}$, bringing the total fluence of this burst to approximately $4.4\times 10^{-4} \rm ~erg~cm^{-2}$ (20-1000 keV) \cite{palshin12}.

\subsection{GRB 090417B (Continuous)}

Detected by \swift, this burst has four broad, overlapping peaks spanning 2130\,s in the BAT 15-150 keV band.  This duration, however, is an upper-limit as observations were stopped because of an earth limb constraint \cite{holland10}.  Apart from its long duration, this burst is an example of a typical GRB, with a PL spectrum in the BAT band (fluence $> 8.2^{+1.0}_{-2.1} \times 10^{-6} \rm ~erg~cm^{-2}$) and a doubly broken poweraw X-ray afterglow with some flaring activity.  The $E_{\rm peak}$ is estimated as $> 150$ keV from the BAT data.  This burst is also dark in UV, optical, and infrared wavelengths with line-of-sight dust extinction as the likely cause.  A host galaxy candidate has been identified and the presumptive redshift for this burst is $z = 0.345$, giving an upper limit of $\gtrsim 6.3\times10^{51}$ erg for the isotropic equivalent energy.

\subsection{GRB 091024A (Interrupted)}

See main text.

\subsection{GRB 100316D}

This local event ($z=0.0591$; Chornock et al. 2010; Starling et al. 2011) shows long lasting and smooth $\gamma$-ray emission, lasting approximately 1300\,s.  The very soft spectrum peaks at $\sim$ 10-42 keV and shows possible evidence of a thermal component in the X-ray emission, similar to GRB 060218A \cite{starling11}.  The fluence of $(5.1 \pm 0.39) \times 10^{-6} \rm ~erg~cm^{-2}$ (15-150 keV) coupled with the low redshift implies an isotropic equivalent energy of approximately $3.9 \times 10^{49}$ erg \cite{starling11}. The optical light curve is also very structured and the late-time emission has been associated with SN emission \cite{wiersema10}.

Recent work by Margutti et al. (2013b) indicates that the thermal component is likely less significant that previously thought and reports a significant soft X-ray excess at late times ($> 10$ days).  In addition, they provide radio observations that infer the kinetic energy coupled to the non-relativistic material of $E_k \sim 10^{52}$ erg and about $10^{49}$ erg coupled with a mildly relativistic ejecta with large opening angle ($> 80^\circ$).  Various theoretical interpretations have been put forward (see \S 5.2) but in general there is growing support that events such as GRB 100316D and 060218A are related to a black hole-torus or magnetar system like `classical' GRBs, but more abundant and significantly less energetic and collimated than their more distant counterparts. 

\subsection{GRB 101225A (Continuous)}

This weak burst detected over various \swift~orbits has both prompt and afterglow emission detected by all three instruments aboard \swift~ and ground-based facilities \cite{racusin11,campana11,thone11}.  Due to gaps in the data, the duration and and fluence are upper-limits of $\gtrsim 1650$\,s and $3 \times 10^{-6} \rm ~erg~cm^{-2}$ in the 15-150 keV band \cite{racusin11,campana11}.  The X-ray light curve is very extended and variable and has a shallow decay followed by a steep decay beginning at $\sim 21$\,ks.  The optical afterglow also shows variability and a gradual decline until the emergence of a likely supernova bump beginning after approximately 10 days \cite{thone11}.

Initially attributed to the accretion of a minor body onto a compact object  at Galactic distances \cite{campana11}, or the merger of a helium star and neutron star at $z \approx 0.33$ \cite{thone11}.  This redshift was obtained by fitting of a supernova template to the late-time optical data.  These scenarios have recently been questioned by the identification of the redshift of this burst at $z = 0.847$ by Levan et al. (2013) from a faint but coincident optical counterpart, which significantly increases the energy budget.  The burst properties (fluence, duration) are upper limits, as the event was already in progress when it entered the \swift~field of view \cite{racusin11} and the true duration could possibly surpass $10^4$\,s \citep{campana11,levan13}.

\subsection{GRB 110709B (Interrupted)}

Similar to GRB 091024A, this burst has two large emission episodes lastin $55.6 \pm 3.2$\,s and $259.2 \pm 8.8$\,s, separated by $\sim 11$ min of quiescence.  This interesting event triggered \swift~ on both of its episodes and lasts about 900\,s in the 15-150 keV band \cite{zhang12}.  A detailed spectral analysis with simultaneous KW data shows hard-to-soft spectral evolution and typical GRB energies across the two pulses: $E_{\rm peak} = 311_{-38}^{+45}$ keV and $116_{-8}^{+9}$ keV, respectively.  This corresponds to $8.95_{-0.62}^{+0.29} \times 10^{-6}$ and $1.34_{-0.07}^{+0.05} \times 10^{-5} \rm ~erg~cm^{-2}$.  This is a an optically dark burst with no redshift determination.  It has been interpreted as a so-called `double-burst' caused by a two-step collapse to a NS then a BH, which creates the two intense emission episodes \cite{zhang12}.

\subsection{GRB 111209A (Continuous)}

This very weak but continuous burst has combined BAT and KW emission lasting $\sim 15$\,ks with $E_{\rm peak} = 310 \pm 53$ corresponding to about $(4.9 \pm 0.61)\times 10^{-4} \rm ~erg~cm^{-2}$.  The redshift has been identified as $z = 0.677$, implying an isotropic equivalent energy of $(5.8 \pm 0.73) \times 10^{53}$ erg \cite{gendre12}.  

The seemingly cannonical X-ray light curve \cite{zhang06} has a shallow decay (slope = $0.544 \pm 0.003$) before breaking at about 20\,ks and exhibiting typical steep decay, plateau, and normal decay behavior \cite{gendre12}.  Although rich in structure and complexity, the X-ray and optical afterglows can be well described within the framework of the external-shock model \cite{stratta13}.  There is late-time rebrightening in the optical light curve that is possibly indicative of supernova emission, but this remains inconclusive.  In the literature, this burst has been interpreted as the collapse of a blue supergiant star with a possible binary companion \cite{gendre12,stratta13}. 

\end{appendix}

\clearpage

\LongTables

\begin{deluxetable}{cccccc}
	 \tablecaption{Cross-Calibrated and Extinction-Corrected Photometry of GRB 091024$^a$}
	 \tablecolumns{6}
  	\tablewidth{0pc}
  	\tablehead{
		 \colhead{$\Delta t^b$} &
		 \colhead{Exposure} &
		 \colhead{Filter} &
		 \colhead{Flux$^c$} &
		 \colhead{Magnitude$^c$} &
		 \colhead{Telescope} 
		 \\
		 \colhead{(s)} &
		 \colhead{(s)} &
		 &
		 \colhead{(mJy)} &
		 &
	}
 	\startdata

440 & 10 & $B$ & 4.35 $\pm$ 0.76 & 14.96 $\pm$ 0.19 & FTN \\
757.2 & 30 & $B$ & 2.18 $\pm$ 0.28 & 15.71 $\pm$ 0.14 & FTN \\
1316 & 60 & $B$ & 1.41 $\pm$ 0.25 & 16.18 $\pm$ 0.19 & FTN \\
2164 & 120 & $B$ & 1.90 $\pm$ 0.18 & 15.86 $\pm$ 0.1 & FTN \\
3260 & 180 & $B$ & 1.65 $\pm$ 0.09 & 16.01 $\pm$ 0.06 & FTN \\
4110 & 30 & $B$ & 1.95 $\pm$ 0.31 & 15.83 $\pm$ 0.17 & FTN \\
4525 & 60 & $B$ & 1.45 $\pm$ 0.17 & 16.15 $\pm$ 0.13 & FTN \\
5336 & 120 & $B$ & 1.09 $\pm$ 0.11 & 16.46 $\pm$ 0.11 & FTN \\
6389 & 180 & $B$ & 0.89 $\pm$ 0.12 & 16.68 $\pm$ 0.15 & FTN \\
7564 & 120 & $B$ & 0.68 $\pm$ 0.16 & 16.98 $\pm$ 0.25 & FTN \\
9429 & 220 & $B$ & 0.47 $\pm$ 0.12 & 17.38 $\pm$ 0.28 & FTN \\
1.11E+04 & 310 & $B$ & 0.46 $\pm$ 0.09 & 17.41 $\pm$ 0.22 & FTN \\
\\
507.8 & 10 & $V$ & 5.53 $\pm$ 0.51 & 14.55 $\pm$ 0.1 & FTN \\
8721 & 10 & $V$ & 1.90 $\pm$ 0.37 & 15.71 $\pm$ 0.21 & FTN \\
3338 & 300 & $V$ & 2.30 $\pm$ 0.24 & 15.5 $\pm$ 0.112 & SRO \\
3654 & 300 & $V$ & 1.86 $\pm$ 0.21 & 15.73 $\pm$ 0.121 & SRO \\
3964 & 300 & $V$ & 1.83 $\pm$ 0.22 & 15.75 $\pm$ 0.133 & SRO \\
4273 & 300 & $V$ & 1.81 $\pm$ 0.21 & 15.76 $\pm$ 0.126 & SRO \\
4582 & 300 & $V$ & 1.97 $\pm$ 0.16 & 15.67 $\pm$ 0.091 & SRO \\
4891 & 300 & $V$ & 1.79 $\pm$ 0.25 & 15.77 $\pm$ 0.151 & SRO \\
5510 & 300 & $V$ & 2.10 $\pm$ 0.21 & 15.6 $\pm$ 0.107 & SRO \\
5819 & 300 & $V$ & 1.87 $\pm$ 0.21 & 15.72 $\pm$ 0.123 & SRO \\
6128 & 300 & $V$ & 1.79 $\pm$ 0.27 & 15.77 $\pm$ 0.163 & SRO \\
230 & 60 & $V$ & 0.98 $\pm$ 0.42 & 16.43 $\pm$ 0.45 & KAIT \\
425 & 20 & $V$ & 8.14 $\pm$ 1.13 & 14.13 $\pm$ 0.15 & KAIT \\
522 & 20 & $V$ & 3.65 $\pm$ 0.99 & 15 $\pm$ 0.29 & KAIT \\
665.5 & 40 & $V$ & 2.65 $\pm$ 0.44 & 15.35 $\pm$ 0.18 & KAIT \\
910.6 & 60 & $V$ & 3.33 $\pm$ 0.43 & 15.1 $\pm$ 0.14 & KAIT \\
\\
201.2 & 10 & $R$ & 1.81 $\pm$ 0.13 & 15.58 $\pm$ 0.08 & FTN \\
231.8 & 10 & $R$ & 2.63 $\pm$ 0.14 & 15.17 $\pm$ 0.06 & FTN \\
264.8 & 10 & $R$ & 3.20 $\pm$ 0.14 & 14.96 $\pm$ 0.05 & FTN \\
844.8 & 30 & $R$ & 4.29 $\pm$ 0.28 & 14.64 $\pm$ 0.07 & FTN \\
1438 & 60 & $R$ & 2.52 $\pm$ 0.19 & 15.22 $\pm$ 0.08 & FTN \\
2339 & 120 & $R$ & 3.47 $\pm$ 0.19 & 14.87 $\pm$ 0.06 & FTN \\
3511 & 180 & $R$ & 2.73 $\pm$ 0.176 & 15.13 $\pm$ 0.07 & FTN \\
4204 & 30 & $R$ & 2.47 $\pm$ 0.20 & 15.24 $\pm$ 0.09 & FTN \\
4650 & 60 & $R$ & 2.17 $\pm$ 0.18 & 15.38 $\pm$ 0.09 & FTN \\
5521 & 120 & $R$ & 2.27 $\pm$ 0.17 & 15.33 $\pm$ 0.08 & FTN \\
6631 & 180 & $R$ & 1.84 $\pm$ 0.15 & 15.56 $\pm$ 0.09 & FTN \\
7564 & 120 & $R$ & 1.42 $\pm$ 0.13 & 15.84 $\pm$ 0.1 & FTN \\
8463 & 30 & $R$ & 1.33 $\pm$ 0.14 & 15.91 $\pm$ 0.11 & FTN \\
9015 & 30 & $R$ & 1.02 $\pm$ 0.13 & 16.2 $\pm$ 0.14 & FTN \\
9378 & 60 & $R$ & 0.98 $\pm$ 0.14 & 16.24 $\pm$ 0.15 & FTN \\
9962 & 120 & $R$ & 1.05 $\pm$ 0.097 & 16.17 $\pm$ 0.1 & FTN \\
1.09E+04 & 180 & $R$ & 0.93 $\pm$ 0.086 & 16.3 $\pm$ 0.1 & FTN \\
1.19E+04 & 120 & $R$ & 0.68 $\pm$ 0.069 & 16.64 $\pm$ 0.11 & FTN \\
1.24E+04 & 30 & $R$ & 0.80 $\pm$ 0.11 & 16.46 $\pm$ 0.15 & FTN \\
4.26E+04 & 1800 & $R$ & 0.11 $\pm$ 0.0099 & 18.65 $\pm$ 0.1 & LT \\
5.46E+04 & 3600 & $R$ & 0.073 $\pm$ 0.0068 & 19.06 $\pm$ 0.1 & LT \\
6.43E+04 & 1800 & $R$ & 0.041 $\pm$ 0.0077 & 19.68 $\pm$ 0.2 & LT \\
642.9 & 180 & $R$ & 5.94 $\pm$ 0.17 & 14.29 $\pm$ 0.031 & SRO \\
839.1 & 180 & $R$ & 4.38 $\pm$ 0.15 & 14.62 $\pm$ 0.038 & SRO \\
1028 & 180 & $R$ & 3.21 $\pm$ 0.16 & 14.95 $\pm$ 0.053 & SRO \\
1219 & 180 & $R$ & 2.73 $\pm$ 0.12 & 15.13 $\pm$ 0.048 & SRO \\
1408 & 180 & $R$ & 2.43 $\pm$ 0.13 & 15.26 $\pm$ 0.058 & SRO \\
1723 & 300 & $R$ & 2.76 $\pm$ 0.12 & 15.12 $\pm$ 0.047 & SRO \\
2039 & 300 & $R$ & 3.51 $\pm$ 0.15 & 14.86 $\pm$ 0.047 & SRO \\
2348 & 300 & $R$ & 3.62 $\pm$ 0.11 & 14.82 $\pm$ 0.034 & SRO \\
2657 & 300 & $R$ & 3.89 $\pm$ 0.14 & 14.75 $\pm$ 0.038 & SRO \\
2966 & 300 & $R$ & 3.46 $\pm$ 0.14 & 14.87 $\pm$ 0.045 & SRO \\
99.42 & 20 & $R$ & 0.52 $\pm$ 0.18 & 16.93 $\pm$ 0.36 & KAIT \\
200.8 & 20 & $R$ & 2.04 $\pm$ 0.13 & 15.45 $\pm$ 0.07 & KAIT \\
307 & 20 & $R$ & 4.17 $\pm$ 0.15 & 14.67 $\pm$ 0.04 & KAIT \\
404 & 20 & $R$ & 6.86 $\pm$ 0.44 & 14.13 $\pm$ 0.07 & KAIT \\
499 & 20 & $R$ & 7.12 $\pm$ 0.26 & 14.09 $\pm$ 0.04 & KAIT \\
594 & 20 & $R$ & 5.66 $\pm$ 0.16 & 14.34 $\pm$ 0.03 & KAIT \\
691 & 20 & $R$ & 5.45 $\pm$ 0.15 & 14.38 $\pm$ 0.03 & KAIT \\
790 & 20 & $R$ & 4.49 $\pm$ 0.17 & 14.59 $\pm$ 0.04 & KAIT \\
889 & 20 & $R$ & 3.81 $\pm$ 0.14 & 14.77 $\pm$ 0.04 & KAIT \\
998 & 20 & $R$ & 3.57 $\pm$ 0.16 & 14.84 $\pm$ 0.05 & KAIT \\
106 & 50 & $R$ & 0.516 $\pm$ 0.27 & 16.94 $\pm$ 0.55 & S-LOTIS \\
167 & 40 & $R$ & 1.34 $\pm$ 0.26 & 15.9 $\pm$ 0.21 & S-LOTIS \\
207.3 & 20 & $R$ & 2.13 $\pm$ 0.20 & 15.4 $\pm$ 0.1 & S-LOTIS \\
234 & 20 & $R$ & 2.61 $\pm$ 0.26 & 15.18 $\pm$ 0.11 & S-LOTIS \\
261.3 & 20 & $R$ & 3.45 $\pm$ 0.19 & 14.87 $\pm$ 0.06 & S-LOTIS \\
308.5 & 60 & $R$ & 4.17 $\pm$ 0.23 & 14.67 $\pm$ 0.06 & S-LOTIS \\
375.3 & 60 & $R$ & 5.56 $\pm$ 0.15 & 14.36 $\pm$ 0.03 & S-LOTIS \\
442 & 60 & $R$ & 7.06 $\pm$ 0.13 & 14.1 $\pm$ 0.02 & S-LOTIS \\
509.2 & 60 & $R$ & 6.93 $\pm$ 0.19 & 14.12 $\pm$ 0.03 & S-LOTIS \\
575.9 & 60 & $R$ & 5.76 $\pm$ 0.16 & 14.32 $\pm$ 0.03 & S-LOTIS \\
643.1 & 60 & $R$ & 5.61 $\pm$ 0.15 & 14.35 $\pm$ 0.03 & S-LOTIS \\
709.9 & 60 & $R$ & 4.97 $\pm$ 0.14 & 14.48 $\pm$ 0.03 & S-LOTIS \\
777.1 & 60 & $R$ & 4.41 $\pm$ 0.16 & 14.61 $\pm$ 0.04 & S-LOTIS \\
844 & 60 & $R$ & 4.06 $\pm$ 0.15 & 14.7 $\pm$ 0.04 & S-LOTIS \\
911.2 & 60 & $R$ & 3.54 $\pm$ 0.16 & 14.85 $\pm$ 0.05 & S-LOTIS \\
977.8 & 60 & $R$ & 3.26 $\pm$ 0.15 & 14.94 $\pm$ 0.05 & S-LOTIS \\
1078 & 120 & $R$ & 2.71 $\pm$ 0.075 & 15.14 $\pm$ 0.03 & S-LOTIS \\
1245 & 180 & $R$ & 2.52 $\pm$ 0.19 & 15.22 $\pm$ 0.08 & S-LOTIS \\
1412 & 120 & $R$ & 2.38 $\pm$ 0.11 & 15.28 $\pm$ 0.05 & S-LOTIS \\
1516 & 120 & $R$ & 2.49 $\pm$ 0.14 & 15.23 $\pm$ 0.06 & S-LOTIS \\
1713 & 180 & $R$ & 2.54 $\pm$ 0.12 & 15.21 $\pm$ 0.05 & S-LOTIS \\
1880 & 120 & $R$ & 3.14 $\pm$ 0.14 & 14.98 $\pm$ 0.05 & S-LOTIS \\
2014 & 120 & $R$ & 3.41 $\pm$ 0.16 & 14.89 $\pm$ 0.05 & S-LOTIS \\
2147 & 120 & $R$ & 3.26 $\pm$ 0.15 & 14.94 $\pm$ 0.05 & S-LOTIS \\
2248 & 60 & $R$ & 3.50 $\pm$ 0.26 & 14.86 $\pm$ 0.08 & S-LOTIS \\
2.46E+05 & 900 & $R$ & 0.0069 $\pm$ 0.00078 & 21.62 $\pm$ 0.12 & Gemini \\
\\
594.2 & 10 & $I$ & 7.10 $\pm$ 0.33 & 13.89 $\pm$ 0.05 & FTN \\
1014 & 30 & $I$ & 4.24 $\pm$ 0.13 & 14.45 $\pm$ 0.03 & FTN \\
1632 & 60 & $I$ & 3.53 $\pm$ 0.16 & 14.65 $\pm$ 0.05 & FTN \\
2749 & 120 & $I $& 4.78 $\pm$ 0.13 & 14.32 $\pm$ 0.03 & FTN \\
4307 & 30 & $I$ & 2.99 $\pm$ 0.14 & 14.83 $\pm$ 0.05 & FTN \\
4854 & 60 & $I$ & 2.83 $\pm$ 0.10 & 14.89 $\pm$ 0.04 & FTN \\
5776 & 120 & $I$ & 2.85 $\pm$ 0.079 & 14.88 $\pm$ 0.03 & FTN \\
6977 & 180 & $I$ & 2.11 $\pm$ 0.097 & 15.21 $\pm$ 0.05 & FTN \\
8802 & 10 & $I$ & 1.50 $\pm$ 0.14 & 15.58 $\pm$ 0.1 & FTN \\
9124 & 30 & $I$ & 1.58 $\pm$ 0.073 & 15.52 $\pm$ 0.05 & FTN \\
9530 & 60 & $I$ & 1.23 $\pm$ 0.068 & 15.79 $\pm$ 0.06 & FTN \\
1.02E+04 & 120 & $I$ & 1.11 $\pm$ 0.071 & 15.91 $\pm$ 0.07 & FTN \\
1.12E+04 & 180 & $I$ & 0.98 $\pm$ 0.045 & 16.04 $\pm$ 0.05 & FTN \\
4.82E+04 & 3600 & $I$ & 0.13 $\pm$ 0.0099 & 18.2 $\pm$ 0.08 & LT \\
5.89E+04 & 1800 & $I$ & 0.076 $\pm$ 0.0070 & 18.82 $\pm$ 0.1 & LT \\
6877 & 180 & $I$ & 2.53 $\pm$ 0.17 & 15.01 $\pm$ 0.074 & SRO \\
174 & 20 & $I$ & 2.01 $\pm$ 0.28 & 15.26 $\pm$ 0.15 & KAIT \\
274 & 20 & $I$ & 4.56 $\pm$ 0.29 & 14.37 $\pm$ 0.07 & KAIT \\
371 & 20 & $I$ & 7.72 $\pm$ 0.50 & 13.8 $\pm$ 0.07 & KAIT \\
468 & 20 & $I$ & 9.80 $\pm$ 0.45 & 13.54 $\pm$ 0.05 & KAIT \\
563 & 20 & $I$ & 7.51 $\pm$ 0.35 & 13.83 $\pm$ 0.05 & KAIT \\
658 & 20 & $I$ & 7.17 $\pm$ 0.26 & 13.88 $\pm$ 0.04 & KAIT \\
757 & 20 & $I$ & 5.96 $\pm$ 0.33 & 14.08 $\pm$ 0.06 & KAIT \\
856 & 20 & $I$ & 5.14 $\pm$ 0.28 & 14.24 $\pm$ 0.06 & KAIT \\
955 & 20 & $I$ & 4.52 $\pm$ 0.33 & 14.38 $\pm$ 0.08 & KAIT \\
2.46E+05 & 900 & $I$ & 0.012 $\pm$ 0.0011 & 20.8 $\pm$ 0.1 & Gemini \\
1.97E+06 & 1800 & $I$ & $>$0.00086 & $>$23.68 & Gemini \\

	\enddata
	\tablenotetext{a}{Frames with similar filters (e.g., $R_C$, $R$) have been calibrated with respect to a common set of standard stars in the reference filter listed. Data were taken with the Faulkes Telescope North (FTN), the Sonoita Research Observatory Telescope (SRO), the Katzman Automatic Imaging Telescope (KAIT), Super-LOTIS (S-LOTIS), the Liverpool Telescope (LT), and Gemini.  See \S 2.2 for further details. }
	\tablenotetext{b}{$\Delta t$ is the midpoint of the exposure in time elapsed (s) since the \swift-BAT trigger.}
   	\tablenotetext{c}{Magnitudes and fluxes have been corrected for Galactic absorption using $E(B-V)=0.98$ mag, which corresponds to $A_B=4.24$, $A_V=3.17$, $A_R=2.58$, and $A_I=1.92$ mag.}
\label{phot2}
\end{deluxetable}

\clearpage

 \begin{deluxetable}{cccccc}
  \tablecaption{Optical Light Curve Multi-Component-Fit Parameters$^a$}
  \tablecolumns{6}
  \tablewidth{0pc}
   \tablehead{ & 
   	\colhead{$\alpha_{\rm rise}$} &
	\colhead{$\alpha_{\rm decay}$} &
	\colhead{$t_{\rm break}^b$} &
	\colhead{$t_{\rm peak}^c$} &
	\colhead{Normalization} \\
          & & & (s) & (s) &
            }
Component 1 & -2.37 $\pm$ 0.13 & 1.83 $\pm$ 0.14 & 450 $\pm$ 19 & 480 $\pm$ 19 &13 $\pm$ 0.19 \\
Component 2 & -4.17 $\pm$ 0.92 & 1.57 $\pm$ 0.57 & 2200 $\pm$ 220 & 2600 $\pm$ 220 & 5.7 $\pm$ 0.84 \\
Component 3 & -15.29 $\pm$ 12.7 & 1.48 $\pm$ 0.81 & 5080 $\pm$ 380 & 5800 $\pm$ 380 & 1.0 $\pm$ 0.54
 \tablenotetext{a}{The reduced $\chi^2$ for the fit is 1.43 for 75 d.o.f. and is performed on the $R$ and $I$ photometry. See Figure 5 and \S 2.3 for details.}
 \tablenotetext{b}{Break time of the Beuermann function fit of each component.}
 \tablenotetext{c}{Time of the light curve peak.}
\label{fits}
\end{deluxetable}

 \begin{deluxetable}{ccccc}
  \tablecaption{Keck/LRIS Absorption-Line Identifications}
  \tablecolumns{5}
  \tablewidth{0pc}
  \tablehead{\colhead{Observed Wavelength} & \colhead{Identification} &
             \colhead{Vacuum Wavelength\tablenotemark{a}} &
             \colhead{Rest-Frame Equivalent Width} \\
             \colhead{(\AA)} & & \colhead{(\AA)} & \colhead{(\AA)}
            }
  \startdata
     $4906.08 \pm 0.40$ & \ion{Fe}{2} & $2344.704$ & $1.43 \pm 0.17$ \\
     $4968.85 \pm 0.55$ & \ion{Fe}{2} & $2374.461$ & $1.02 \pm 0.32$ \\
     $4985.74 \pm 0.40$ & \ion{Fe}{2} & $2382.765$ & $2.11 \pm 0.11$ \\
     $5412.51 \pm 0.40$ & \ion{Fe}{2} & $2586.650$ & $1.67 \pm 0.05$ \\
     $5440.37 \pm 0.40$ & \ion{Fe}{2} & $2600.173$ & $2.05 \pm 0.12$ \\
     $5851.19 \pm 0.49$ & \ion{Mg}{2} & $2796.352$ & $3.02 \pm 0.25$ \\
     $5866.18 \pm 0.42$ & \ion{Mg}{2} & $2803.531$ & $2.32 \pm 0.35$ \\
     $5968.43 \pm 0.30$ & \ion{Mg}{1} & $2852.964$ & $1.32 \pm 0.18$ \\
     $8232.54 \pm 0.20$ & \ion{Ca}{2} & $3934.777$ & $1.73 \pm 0.03$ \\
     $8305.99 \pm 0.20$ & \ion{Ca}{2} & $3969.591$ & $1.17 \pm 0.03$
  \enddata
  \tablenotetext{a}{Ref: Morton (1991).}
\label{lris}
\end{deluxetable}

\begin{deluxetable}{cccccccc}
  \tablecaption{Konus-Wind Time-Integrated Spectral Parameters for Each $\gamma$-ray Emission Episode$^a$}
  \tablecolumns{8}
  \tablewidth{0pc}
  \tablehead{
  	   \colhead{GBM time interval$^b$} & 
	   \colhead{KW time interval$^c$} &
	   \colhead{$\Delta t$} & 
             \colhead{$\alpha_{\rm CPL}$} &
             \colhead{$E_{\rm peak,CPL}$} &
             \colhead{$\alpha_{\rm GRBM}$} &
             \colhead{$E_{\rm peak,GRBM}$} &
             \colhead{$\beta_{\rm GRBM}$}  \\ 
             \colhead{(s)} & 
             \colhead{(s)} &
             \colhead{(s)} & 
             & 
             \colhead{(keV)} & 
             &
             \colhead{(keV)} &
            } 
  \startdata
     (-3.8, 67.8) & (-6.6, 65.0) & 71.6  & $-1.07^{+0.09}_{-0.08}$ & 495$^{+106}_{-69}$ & $-1.06^{+0.10}_{-0.08}$ & 474$^{+111}_{-74}$ & -2.5 \\
     (622.7, 664.7) & (619.9, 661.9) & 42  & $-1.59^{+0.10}_{-0.11}$ & 374$^{+490}_{-128}$ & $-1.58^{+0.13}_{-0.10}$ & 495$^{+494}_{-140}$ & -2.5   \\
     (838.8, 1070.2) & (836.0, 1067.4) & 231.4  & $-1.42^{+0.4}_{-0.4}$ & 246$^{+26}_{-26}$ & $-1.38^{+0.06}_{-0.05}$ & 216$^{+31}_{-25}$ & -2.5 \\
     \\
     (-6.1, 82.2) & (-8.9, 79.4) & 88.3 & $-1.10^{+0.09}_{-0.08}$ & $527^{+125}_{-79}$ &   $-1.09^{+0.10}_{-0.08}$ & $508^{+130}_{-84}$ & -2.5 \\
     (612.1, 718.1) & (609.4, 715.3) & 105.9 & $-1.61^{+0.17}_{-0.13}$ & $184^{+137}_{-48}$ & $-1.57^{+1.96}_{-0.17}$ & $161^{+148}_{-87}$ & -2.5 \\
     (835.1, 1312.8) & (833.1, 1310.0) & 476.9 & $-1.49^{+0.04}_{-0.05}$ & $ 255^{+41}_{-28}$ & $-1.46^{+0.06}_{-0.06}$ & $230^{+46}_{-34}$ & -2.5 \\
     \\
     \\
     (82.2, 606.6) & (79.4, 609.4) & 524.4  & $-1.73^{+0.13}_{-0.12}$ & - & - & - & - 
  \enddata
   \tablenotetext{a}{The first two sets of parameters are fits over the approximate $Fermi$/GBM durations and the latter sets are fits using the Bayesian Block derived durations.  Each time interval is fit with a cutoff power law (CPL) and GRBM model with $\beta_{\rm GRBM} = -2.5$.  The last line is a simple power law fit over the interval between the first and second emission episodes. Errors are approximated with the bootstrap method and given to the $1\sigma$ confidence level.}
  \tablenotetext{b}{Seconds with respect to the first \fermi-GBM trigger of GRB 091024A.}
   \tablenotetext{c}{Seconds with respect to the Konus-Wind trigger.}
\label{spectra}
\end{deluxetable}

\clearpage

\begin{deluxetable}{cccccccc}
  \tablecaption{Time-Resolved Spectral Parameters for Joint Fits of First Emission Episode$^a$}
  \tablecolumns{8}
  \tablewidth{0pc}
  \tablehead{
  	    \colhead{Time interval$^b$} & \colhead{Model} &
             \colhead{$\alpha$} &
             \colhead{$E_{\rm peak}$}
             & \colhead{$\chi^2/$d.o.f.} &  \colhead{Instrument} \\  \colhead{(s)}
             & & & \colhead{(keV)} & &
            } 
  \startdata
     (-8.887, -2.999) & CPL & $-1.24^{+0.39}_{-0.26}$ & 572 ($>193$) & 78.7/57 & KW+BAT \\
     (-2.999, 8.777) & CPL & $-0.93^{+0.11}_{-0.11}$ & $465^{+140}_{-91}$ & 55.5/57& KW+BAT \\
     (8.777, 23.497) & CPL & $-1.04^{+0.21}_{-0.19}$ & $505^{+519}_{-179}$ & 36.2/57 & KW+BAT \\
     (23.497, 38.217) & CPL & $-1.07^{+0.56}_{-0.42}$ & 457 ($>193$) & 46.1/57 & KW+BAT \\
     (38.217, 49.993) & CPL & $-1.31^{+0.19}_{-0.16}$ & 783 ($>362$) & 67.1/57 & KW+BAT \\
     (49.993, 76.489) & CPL & $-1.33^{+0.27}_{-0.21}$ & 643 ($>246$) & 56.7/57 & KW+BAT \\
     (-8.887, 76.489) & CPL & $-1.12^{+0.09}_{-0.09}$ & $523^{+198}_{-118}$ & 49/57 & KW+BAT \\
  \enddata
   \tablenotetext{a}{Time-resolved spectral parameters for joint fits on Konus-Wind and \swift-BAT data for the first emission episode of GRB 091024A. Uncertainties are approximated with the bootstrap method and given to the $1\sigma$ confidence level.}
  \tablenotetext{b}{Time intervals derived with a Bayesian Block technique from the Konus-Wind light curve.  Times are relative to the \swift-BAT trigger.}
\label{spectra2}
\end{deluxetable}


\begin{deluxetable}{ccccccccc}
  \tablecaption{Ultra-Long GRB Sample}
  \tablecolumns{9}
  \tablewidth{0pc}
  \tablehead{
  	    \colhead{GRB}&
	    \colhead{Duration$^a$} & 
	    \colhead{Redshift} &
	    \colhead{Fluence} &
	    \colhead{Energy range$^b$} &
	    \colhead{$E_{\rm peak}^c$} &
             \colhead{$E_{\rm iso}^d$} &
             \colhead{Comment} & 
             \colhead{Ref.} 
              \\ 
             & 
             \colhead{(s)} & 
             & 
             \colhead{($10^{-4}$\,erg\,cm$^{-2}$)}&
             \colhead{(keV)} &
             \colhead{(keV)} &
             \colhead{($10^{52}$\,erg)} & 
             & 
            } 
  \startdata
	840304A & 1200 & -- & 28 & -- & -- & $\sim 760$ & 2 broad pulses & 1\\
	& & & & & & & + 1000s extended tail & \\
	971208A & 2500 & -- & 2.55 & 15--1000& $144$ & $\sim 69$ & 1 FRED-like pulse & 2, 3, 4 \\
	020410A & 1550 & $\sim 0.5^e$ & 0.28 & 15--1000 & 900 & $\sim 1.8$ & Multi-episode & 5,6 \\
	060218A & 2100 & 0.0331 & 0.17 & 15--150 & 4.9 & 0.0062 & 1 pulse &  7,8,9 \\
	060814B & 2700 & -- & 2.35 & 18--1170 & 341 & $\sim 64$ & 1 FRED-like pulse & 4 \\
	080407A & 2100 & -- & 4.4 & 20--1000 & 325, 114 & 120 & Multi-episode & 10\\
	090417B & $>2130$ & 0.345 & 0.082 & 15--150 & $>150$ & $>$0.63 & 4 broad peaks & 11\\
	091024A & 1300 & 1.092 & 1.5 (1.7)$^f$ & 10--10$^4$ & 508, 161, 230 & 44 (52)$^g$ & Multi-episode & this work, 12 \\
	100316D & 1300 & 0.0591 & 0.051 & 15-150 & 10--42 & 0.0039 & Overlapping emission & 13 \\
	101225A & 1650-10$^4$ & 0.847 & $>$0.03 & 15--150 & $\sim 20$  & 1.2 & Continuous emission &14,15,16,17 \\
	110709B & 900 & -- & 0.22 & 15--150 & 311, 116 & $\sim 10$ & Multi-episode & 18 \\
	111209A & $\sim$ 13,000$^h$ & 0.677 & 4.9 & 15--150 & 310 & 58 & Continuous emission & 19,20 \\
	 
  \enddata
  \tablenotetext{}{References --- 1: \cite{klebesadel84},
   2: \citep{connaughton97}, 3: \cite{giblin02}, 4: \cite{palshin08}, 
   5: \cite{nicastro04}, 6: \cite{levan05},
   7: \cite{campana06}, 8: \cite{soderberg06}, 9: \cite{liang06a}
   10: \cite{palshin12},
    11: \cite{holland10}, 
    12: \cite{gruber11}, 
    13: \cite{starling11}
    14: \cite{racusin11}, 15: \cite{campana11}, 16: \cite{thone11}, 17:  \cite{levan13}
    18: \cite{zhang12}, 
    19: \cite{gendre12}, 20: \cite{stratta13}.}
\tablenotetext{a}{Approximate duration of the entire $\gamma$-ray interval, including intervals of low-level or quiescent emission.}
\tablenotetext{b}{Energy range in which the fluence was reported.}
\tablenotetext{c}{Lines with multiple entries denote the peak energy of each emission episode.}
\tablenotetext{d}{For bursts with unknown redshift, the radiated isotropic equivalent energy is estimated at $z=1$.  }
\tablenotetext{e}{Levan et al. (2005) find this estimate for the redshift from modeling and fitting of the late-time lightcurve with expected supernova emission.  Nicastro et al. (2004) provide an estimate of 0.9 $< z <$ 1.5 from the Amati relation.}
\tablenotetext{f}{Second value is the total fluence including the interpulse region between the first and second $\gamma$-ray episodes.}
\tablenotetext{g}{Second value includes the fluence contribution from the interpulse region.}
\tablenotetext{h}{We estimate the $\gamma$-ray duration to be consistent in our burst selection.  Gendre et al. (2013) report a duration that includes the longer-lasting X-ray emission attributed to the central engine, increasing their value to $\sim 25$\,ks.}
\label{sample}
\end{deluxetable}



\begin{thebibliography}{}

\bibitem[Akerlof et al. 1999]{akerlof99} Akerlof, C., et al. 1999, Nature, 398, 400
\bibitem[Amati et al. 2002]{amati02} Amati, L., et al. 2002, A\&A, 390 81
\bibitem[Arnold et al. 2013]{arnold12} Arnold, D. M., Steele, I. A., Bates, S. D., Mottram, C. J., Smith, R. J. 2012, SPIE, 8846, 2 
\bibitem[Band et al. 1993]{band93} Band, D., et al. 1993, ApJ, 413, 281
\bibitem[Berger et al. 2003]{berger03} Berger, E., Kulkarni, S. R., Frail, D. A. 2003, ApJ, 590, 379
\bibitem[Bissaldi \& Connaughton 2009]{bissaldi09} Bissaldi, E., Connaughton, V. 2009, GCN Circulars, 10070, 1
\bibitem[Bo$\ddot{\rm e}$r et al. 2006]{boer06} Bo$\ddot{\rm e}$r, M., Atteia, J. L., Damerdji, Y., Gendre, B., Klotz, A., Stratta, G. 2006, ApJ, 638, L71
\bibitem[Bromberg et al. 2011]{bromberg11} Bromberg, O., Nakar, E., Piran, T. 2011, ApJ, 739, 55
\bibitem[Burrows et al. 2005]{burrows05} Burrows, D. N., et al. 2005, Science, 309, 1833
\bibitem[Campana et al. 2006]{campana06} Campana, S., et al. 2006, Nature, 442, 1008
\bibitem[Campana et al. 2011]{campana11} Camapana, S., et al. 2011, Nature, 480, 69
\bibitem[Cano et al. 2009]{cano09} Cano, Z., et al. 2009, GCN Circulars, 10066, 1
\bibitem[Chincarini et al. 2010]{chincarini10} Chincarini, G., et al. 2010, MNRAS, 406, 2113
\bibitem[Chornock et al. 2009]{chornock09} Chornock, R., Li, W., Filippenko, A. V. 2009, GCN Circulars, 10075, 1
\bibitem[Chornock et al. 2010]{chornock10} Chornock, R., et al. 2010, arXiv:1004.2262
\bibitem[Connaughton et al. 1997]{connaughton97} Connaughton, V., et al. 1997, IAUC, 6785, 1
\bibitem[Drago \& Pagliara 2007]{drago07} Drago, A., Pagliara, G. 2007, ApJ, 665, 1227
\bibitem[Drago et al. 2008]{drago08} Drago, A., Pagliara, G., Schaffner-Bielich, J. 2008, J. Ph. G. Nucl. Part. Phys., 35, 014052
\bibitem[Evans et al. 2007]{evans07} Evans, P. A., et al. 2007, A\&A, 469, 379
\bibitem[Evans et al. 2009]{evans09} Evans, P. A., et al. 2009, MNRAS, 397, 1177
\bibitem[Fan \& Wei 2005]{fan06} Fan, Y. Z., Wei, D. M. 2005, MNRAS, 364, L42
\bibitem[Feroci et al. 2001]{feroci01} Feroci, M., et al. 2001, A\&A, 378, 441
\bibitem[Fenimore \& Ramirez-Ruiz 2000]{fenimore00} Fenimore, E. E., \& Ramirez-Ruiz, E. 2000, arXiv:astro-ph/0004176
\bibitem[Filippenko et al. 2001]{filippenko01} Filippenko, A. V., Li, W. D., 
Treffers, R. R., Modjaz, M. 2001, in Small-Telescope Astronomy on Global 
Scales, ed. B. Paczy\'{n}ski, W.-P. Chen, C. Lemme (San Francisco: ASP), 121
\bibitem[Fishman et al. 1989]{fishman89} Fishman, G. J., et al. 1989, in Proc. of the GRO Science Workshop, ed. W. N. Johnson (NASA/GSFC), 39
\bibitem[Freedman \& Waxman 2001]{fandw01} Freedman, D. L., Waxman, E. 2001, ApJ, 547, 922
\bibitem[Frontera et al. 2009]{frontera09} Frontera, F., et al. 2009, ApJS, 180, 192
\bibitem[Gehrels et al. 2004]{gehrels04} Gehrels, N., et al. 2004, ApJ, 611, 1005
\bibitem[Gendre et al. 2010]{gendre10} Gendre, B., et al. 2010, MNRAS, 405, 237
\bibitem[Gendre et al. 2013]{gendre12} Gendre, B., et al. 2013, ApJ, 766, 30
\bibitem[Giblin et al. 2002]{giblin02} Giblin, T. W., et al. 2002, ApJ, 570, 573
\bibitem[Golenetskii et al. 2009]{golenetskii09} Golenetskii, R., Aptekar, R., Mazets, E., Pal'shin, V., Frederiks, D., Oleynik, P., Ulanov, M., Svinkin, D. 2009, GCN Circulars, 10083, 1
\bibitem[Gomboc et al. 2008]{gomboc08} Gomboc, A., et al. 2008, ApJ, 687, 443
\bibitem[Gruber et al. 2011]{gruber11} Gruber, D., et al. 2011, A\&A, 528, A15
\bibitem[Guidorzi et al. 2006a]{guidorzi06} Guidorzi, C., et al. 2006a, PASP, 118, 288
\bibitem[Guidorzi et al. 2006b]{guidorzi06b} Guidorzi, C., et al. 2006b, MNRAS, 371, 843
\bibitem[Guidorzi et al. 2009]{guidorzi09} Guidorzi, C., et al. 2009, A\&A, 499, 439
\bibitem[Guidorzi et al. 2011a]{guidorzi11a} Guidorzi, C., et al. 2011a, MNRAS, 417, 2124
\bibitem[Guidorzi et al. 2011b]{guidorzi11b} Guidorzi, C., et al. 2011b, A\&A, 526, A49
\bibitem[Harrison \& Kobayashi 2013]{harrison12} Harrison, R., Kobayashi, S. 2013, ApJ, 772, 101
\bibitem[Henden et al. 2009]{henden09} Henden, A., Gross, J., Denny, B., Terrell, D., Conney, W. 2009, GCN Circulars, 10073, 1
\bibitem[Holland et al. 2010]{holland10} Holland, S. T. 2010, ApJ, 717, 223
\bibitem[Horne 1986]{h86} Horne, K. PASP, 1986, 98, 609
\bibitem[in 't Zand et al. 2004]{intZand03} in 't Zand, J. J. M., Heise, J., Kippen, R. M., Woods, P. M., Guidorzi, C., Montanari, E., Frontera, F. 2004, ASPC, 312, 18 
\bibitem[Kalberla et al. 2005]{kalberla05} Kalberla, P. M., Burton, W. B., Hartmann, D., Arnal, E. M., Bajaja, E., Morras, R., P$\ddot{\rm o}$ppel, W. G. L. 2005, A\&A, 440, 775
\bibitem[]{} Klebesadel, R. W., Strong, I. B., Olson, R. A. 1973, ApJ, 182, L85
\bibitem[Klebesadel et al. 1984]{klebesadel84} Klebesadel, R. W., Laros, J. G., Fenimore, E. E. 1984, BAAS, 16, 1016
\bibitem[Kobayashi et al. 1997]{kobayashi97} Kobayashi, S., Piran, T., Sari, R. 1997, ApJ, 490, 92
\bibitem[Kobayashi \& Sari 2000]{kobayashi00} Kobayashi, S., Sari, R. 2000, ApJ, 542, 819
\bibitem[Kobayashi \& Zhang 2003]{kobayashi03} Kobayashi, S., Zhang, B. 2003, ApJ, 582, L75
\bibitem[Kopa$\check{\rm c}$ et al. 2013]{kopac13} Kopa$\check{\rm c}$, D., et al. 2013, ApJ, 772, 73
\bibitem[Kouveliotou et al. 1993]{kouveliotou93} Kouveliotou, C., Meegan, C. A., Fishman, G. J., Bhat, N. P., Briggs, M., Koshut, T. M., Paciesas, W. S., Pendelton, G. N. 1993, ApJ, 413, L101
\bibitem[Kulkarni et al. 1999]{kulkarni99} Kulkarni, S., et al. 1999, ApJ, 522, 97
\bibitem[Kumar et al. 2008a]{kumar08a} Kumar, P., Narayan, R., Johnson, J. L.  2008a, MNRAS, 388, 1729 
\bibitem[Kumar et al. 2008b]{kumar08b} Kumar, P, Narayan, R., Johnson, J. L. 2008b, Science, 321, 376
\bibitem[Laskar et al. 2013]{laskar13} Laskar, T., et al. 2013, arXiv:1305.2453
\bibitem[Lazzati et al. 2010]{lazzati10} Lazzati, D., Morsony, B. J., Begelman, M. C. 2009, ApJ, 700, L47
\bibitem[Lazzati et al. 2012]{lazzati12} Lazzati, D., Morsony, B. J., Blackwell, C. H., Begelman, M. C. 2012, ApJ, 750, 68
\bibitem[Levan et al. 2005]{levan05} Levan, A., et al. 2005, ApJ, 624, 880
\bibitem[Levan et al. 2013]{levan13} Levan, A., et al. 2013, arXiv:1302:2352
\bibitem[Li 2001]{li01} Li, T.-P. 2001, Chin. J. Astron. Astrophys., 1, 313 
\bibitem[Li \& Muraki 2001]{limuraki01} Li, T. P.,  Muraki, Y. 2001, ApJ, 578, 374
\bibitem[Li et al. 2003]{wli03}Li, W., Filippenko, A. V., Chornock, R., Jha, S. 2003, PASP, 115, 844.
\bibitem[Liang et al. 2013]{liang12} Liang, E. W., et al. 2013, ApJ, 774, 13
\bibitem[Liang et al. 2006a]{liang06a} Liang, E. W., Zhang, B. B., Stamatikos, M., Zhang, B., Norris, J., Gehrels, N., Zhang, J., Dai, Z. G. 2006a, ApJ, 653, L81
\bibitem[Liang et al. 2006b]{liang06b} Liang, E. W., et al. 2006b, ApJ, 646, 351
\bibitem[Lloyd-Ronning \& Zhang 2004]{lloydronning04} Lloyd-Ronning, N. M., Zhang, B. 2004, ApJ, 613, 477
\bibitem[L\'{o}pez-C\'{a}mara et al. 2010]{lopez10} L\'{o}pez-C\'{a}mara, D., Lee, W. H., Ramirez-Ruiz, E. 2010, ApJ, 716, 1308
\bibitem[Margutti et al. 2011]{margutti11} Margutti, R., Bernardini, G., Barniol-Duran, R., Guidorzi, C., Shen, R. F., Chincarini, G. 2011, MNRAS, 410, 1064
\bibitem[Margutti et al. 2008]{margutti08} Margutti, R., Guidorzi, C., Chincarini, G., Pasotti, F., Covino, S., Mao, J. 2008, arXiv:0809.0189
\bibitem[Margutti et al. 2010]{margutti10} Margutti, R., Guidorzi, C., Chincarini, G., Bernardini, M. G., Genet, F., Mao, J., Pasotti, F. 2010, MNRAS, 406, 2149
\bibitem[Margutti et al. 2013a]{margutti13} Margutti, R., et al. 2013a, MNRAS, 428, 729
\bibitem[Margutti et al. 2013b]{margutti13b} Margutti, R., et al. 2013b, arXiv:1308.1687
\bibitem[Marshall et al. 2009]{marshall09} Marshall, F. E., et al. 2009, GCN Circulars, 10062, 1
\bibitem[Matheson et al. 2000]{mfh+00} Matheson, T., Filippenko, A. V., Ho, L. C., Barth, A. J., Leonard, D. C. 2000, AJ, 120, 1499
\bibitem[Matzner 2003]{matzner03} Matzner, C. D. 2003, MNRAS, 345, 575
\bibitem[Melandri et al. 2008]{melandri08} Melandri, A., et al. 2008, ApJ, 606, 1209
\bibitem[Melandri et al. 2009]{melandri09} Melandri, A., et al. 2009, MNRAS, 395, 1941
\bibitem[Melandri et al. 2010]{melandri10} Melandri, A., et al. 2010, ApJ, 723, 1331
\bibitem[M\'{e}sz$\acute{\rm a}$ros \& Rees 1999]{meszaros99} M\'{e}sz$\acute{\rm a}$ros, P., Rees, M. J. 1999, MNRAS, 306, 39
\bibitem[Mirabal et al. 2006]{mirabal06} Mirabal, N., Halpern, J. P. 2006, GCN Circulars, 4792, 1
\bibitem[Molinari et al. 2007]{molinari07} Molinari, E., et al. 2007, A\&A, 469, 13
\bibitem[Morton 1991]{m91} Morton, D. C. 1991, ApJS, 77, 119
\bibitem[Mundell et al. 2009]{mundell09} Mundell, C. G., et al. 2009, GCN Circulars, 10063, 1
\bibitem[Nakauchi et al. 2012]{nakauchi12} Nakauchi, D., Suwa, Y., Sakamoto, T., Kashiyama, K.,  Nakamura, T. 2012, ApJ, 759, 128
\bibitem[Nakauchi et al. 2013]{nakauchi13} Nakauchi, D., Kashiyama, K., Suwa, Y., Nakamura, T. 2013, arXiv:1307:5061
\bibitem[Nakar \& Piran 2002]{nakar02} Nakar, E., Piran, T. 2002, ApJ, 572, 139
\bibitem[Nakar \& Piran 2002]{nakar04}Nakar, E., Piran, T. 2004, MNRAS, 353, 647
\bibitem[Nakar \& Sari 2012]{nakar12} Nakar, E., Sari, R. 2012, ApJ, 747, 88
\bibitem[Nicastro et al. 2004]{nicastro04} Nicastro, L., et al. 2004, A\&A, 427, 445
\bibitem[Oke et al. 1995]{occ+95} Oke, J. B., et al. 1995, PASP, 107, 375 
\bibitem[]{} Paciesas W. S., et al. 1999, ApJS, 122, 465
\bibitem[Page \& Marshall 2009]{page09} Page, K. L., Marshall, F. E. 2009, GCN Circulars, 10069, 1
\bibitem[Page et al. 2007]{page07} Page, K. L., et al. 2007, ApJ, 663, 1125
\bibitem[Pal'shin et al. 2008]{palshin08} Pal'shin, V., et al. 2008, AIPC, 1000, 117
\bibitem[Pal'shin et al. 2012]{palshin12} Pal'shin, V., et al. 2012, Proc. Science, GRB 2012, 40, 2012
\bibitem[Panaitescu \& Kumar 2001]{pandk01} Panaitescu, A., Kumar, P. 2001, ApJ, 560, L49
\bibitem[Park et al. 1997]{park97} Park, H. S. 1997, astro-ph/9711170
\bibitem[Park et al. 2002]{park02} Park, H. S., et al. 2002, ApJ, 571, 131
\bibitem[Peng et al. 2013]{peng13} Peng, F. K., Hu, Y. D., Xi, S. Q., Wang, X. G, Lu, R. J., Liang, E. W., Zhang, B. 2013, arXiv:1302:4876
\bibitem[Podsiadlowski 2007]{podsiadlowski07} Podsiadlowski, P. 2007, RSPTA, 365, 1163 
\bibitem[Podsiadlowski 2010]{podsiadlowski10} Podsiadlowski, P. Ivanova, N., Justham, S., Rappaport, S. 2010, MNRAS, 406, 840
\bibitem[Qin et al. 2013]{qin13} Qin, Y., et al. 2013, ApJ, 763, 15
\bibitem[Quilligan et al. 2002]{quilligan02} Quilligan, F., McBreen, B., Hanlon, L., McBreen, S., Hurley, K. J., Watson, D. 2002, A\&A, 385, 377
\bibitem[Racusin et al. 2008]{racusin08} Racusin, J., et al. 2008, Nature, 455, 183
\bibitem[Racusin et al. 2011]{racusin11} Racusin, J., Cummings, J., Holland, S., Krimm, H., Oates, S. R., Page, K., Siegel, M. 2011, GCN Report, 314.1, 1
\bibitem[Ramirez-Ruiz, Merloni, Rees]{ramirez01} Ramirez-Ruiz, E., Merloni, A., Rees, M. J. 2001, MNRAS, 324, 1147
\bibitem[Reichart et al. 2001]{reichart01} Reichart, D. E., et al. 2001, ApJ, 552, 57
\bibitem[Sakamoto et al. 2011]{sakamoto11} Sakamoto, T., et al. 2011, ApJS, 195, 2
\bibitem[Sari \& Piran 1999]{sari99} Sari, R., Piran, T. 1999, ApJ, 520, 641
\bibitem[Soderberg et al. 2006]{soderberg06} Soderberg, A., et al. 2006, Nature, 442, 1014
\bibitem[Starling et al. 2011]{starling11} Starling, R. L. C., et al. 2011, MNRAS, 411, 2792
\bibitem[Steele et al. 2009]{steele09} Steele, I. A., Mundell, C. G., Smith, R. J., Kobayashi, S., Guidorzi, C. 2009, Nature, 462, 767
\bibitem[Stratta et al. 2013]{stratta13} Stratta, G., et al. 2013, arXiv:1306.1699
\bibitem[Suwa \& Ioka, 2011]{suwa11} Suwa, Y, Ioka, K. 2011, ApJ, 726, 107
\bibitem[Th$\ddot{\rm o}$ne et al. 2011]{thone11} Th$\ddot{\rm o}$ne, C. C., et al. 2011, Nature, 480, 72
\bibitem[Tikhomirova \& Stern 2005]{tikhomirova05} Tikhomirova, Y. Y., Stern, B. E. 2005, AstL, 31, 291
\bibitem[Updike et al. 2009]{updike09} Updike, A. C., Hartmann, D. H., Milne, P. A., Williams, G. G. 2009, GCN Circulars, 10074, 1
\bibitem[van Dokkum 2001]{v01} van Dokkum, P. G. 2001, PASP, 113, 1420
\bibitem[Vestrand et al. 2005]{vestrand05} Vestrand, W. T., et al. 2005, Nature, 435, 178
\bibitem[Vestrand et al. 2006]{vestrand06} Vestrand, W. T., et al. 2006, Nature, 442, 172
\bibitem[Virgili et al. 2012]{virgili12} Virgili, F. J., Qin, Y., Zhang, B., Liang, E. W. 2012, MNRAS, 424, 2821
\bibitem[Wade \& Horne 1988]{wh88} Wade, R. A., Horne, K. 1988, ApJ, 324, 411
\bibitem[Wei et al. 2006]{wei06} Wei, D. M., Yan, T., Fan, Y. Z. 2006, ApJ, 636, L69
\bibitem[Wiersema et al. 2010]{wiersema10} Wiersema K., DÕAvanzo P., Levan A. J., Tanvir N. R., Malesani D., Covino S. 2010, GCN Circulars, 10525
\bibitem[Woosley 2011]{woosley11} Woosley, S. E. 2010, ``Gamma-Ray Bursts'' (Cambridge: Cambridge University Press) (available online: arXiv:1105.4193)
\bibitem[Woosley \& Heger 2012]{woosley12} Woosley, S. E., Heger, A., 2012, ApJ, 752, 32
\bibitem[Wu et al. 2013]{wu13} Wu, X. F., Hou, S. J., Lei, W. H. 2013, ApJL, 767, L36
\bibitem[Yamazaki et al. 2006]{yamazaki06} Yamazaki, R., Toka, K., Ioka, K., Nakamura, T. 2006, MNRAS, 369, 311
\bibitem[Zhang et al. 2003]{zhang03} Zhang, B., Kobayashi, S., M\'{e}sz$\acute{\rm a}$ros, P. 2003, ApJ, 595, 950
\bibitem[Zhang \& Kobayashi 2005]{zhang05} Zhang, B., Kobayashi, S. 2005, ApJ, 628, 315
\bibitem[Zhang et al. 2006]{zhang06} Zhang, B., et al. 2006, ApJ, 642, 354
\bibitem[Zhang et al. 2007]{zhang07a} Zhang, B., et al. 2007, ApJ, 655, 989
\bibitem[Zhang et al. 2009a]{zhang09} Zhang, B., et al. 2009a, ApJ, 703, 1696
\bibitem[Zhang et al. 2009b]{zhangbb09} Zhang, B. B., Zhang, B., Liang, E. W., Wang, X. Y. 2009b, ApJ, 690, L10
\bibitem[Zhang et al. 2012]{zhang12} Zhang, B. B., et al. 2012, ApJ, 748, 132

\bibitem[]{}

\end{thebibliography}
\end{document}